\documentclass[aps,prc,twocolumn,showpacs,superscriptaddress,nofootinbib,longbibliography]{revtex4-1}

\usepackage{graphicx}
\usepackage{dcolumn}
\usepackage{bm}
\usepackage{epstopdf}
\usepackage{multirow}
\usepackage{enumitem}
\usepackage{paralist}
\usepackage{booktabs}
\usepackage[version=3]{mhchem}
\usepackage{color}
\usepackage{placeins}
\usepackage{verbatim}

\begin{document}

\preprint{APS/123-QED}

\title{High-resolution laser spectroscopy of $^\text{27-32}$Al}

\author{H.~Heylen}
\email{Corresponding author: hanne.heylen@cern.ch}
\affiliation{Experimental Physics Department, CERN, CH-1211 Geneva 23, Switzerland}
\affiliation{KU Leuven, Instituut voor Kern- en Stralingsfysica, 3001 Leuven, Belgium}
\affiliation{Max-Planck-Institut f\"ur Kernphysik, D-69117 Heidelberg, Germany}
\author{C.S.~Devlin}
\affiliation{Oliver Lodge Laboratory, Oxford Street, University of Liverpool, Liverpool, L69 7ZE, United Kingdom}
\author{W.~Gins}
\affiliation{KU Leuven, Instituut voor Kern- en Stralingsfysica, 3001 Leuven, Belgium}
\affiliation{Department of Physics, University of Jyv\"askyl\"a, PB 35 (YFL), 40014 Jyv\"askyl\"a, Finland}
\author{M.L. Bissell}
\affiliation{School of Physics and Astronomy, The University of Manchester, Manchester, M13 9PL, United Kingdom}
\author{K. Blaum}
\affiliation{Max-Planck-Institut f\"ur Kernphysik, D-69117 Heidelberg, Germany}
\author{B. Cheal}
\affiliation{Oliver Lodge Laboratory, Oxford Street, University of Liverpool, Liverpool, L69 7ZE, United Kingdom}
\author{L. Filippin}
\affiliation{Spectroscopy, Quantum Chemistry and Atmospheric Remote Sensing (SQUARES), Universit\'e libre de Bruxelles, 1050 Brussels, Belgium}
\author{R.F. Garcia Ruiz}
\altaffiliation[Present address: ]{Massachusetts Institute of Technology, Cambridge, MA, USA}
\affiliation{Experimental Physics Department, CERN, CH-1211 Geneva 23, Switzerland}
\affiliation{School of Physics and Astronomy, The University of Manchester, Manchester, M13 9PL, United Kingdom}
\author{\mbox{M. Godefroid}}
\affiliation{Spectroscopy, Quantum Chemistry and Atmospheric Remote Sensing (SQUARES), Universit\'e libre de Bruxelles, 1050 Brussels, Belgium}
\author{C. Gorges}
\affiliation{Institut f\"ur Kernphysik, Technische Universit\"at Darmstadt, D-64289 Darmstadt, Germany}
\author{J. D. Holt}
\affiliation{TRIUMF, 4004 Wesbrook Mall, Vancouver, British Columbia, V6T 2A3, Canada}
\affiliation{Department of Physics, McGill University, 3600 Rue University, Montr\'eal, QC H3A 2T8, Canada}
\author{A. Kanellakopoulos}
\affiliation{KU Leuven, Instituut voor Kern- en Stralingsfysica, 3001 Leuven, Belgium}
\author{S. Kaufmann}
\affiliation{Institut f\"ur Kernphysik, Technische Universit\"at Darmstadt, D-64289 Darmstadt, Germany}
\affiliation{Institut f\"ur Kernchemie, Universit\"at Mainz, D-55128 Mainz, Germany}
\author{\'A. Koszor\'us}
\affiliation{KU Leuven, Instituut voor Kern- en Stralingsfysica, 3001 Leuven, Belgium}

\author{\mbox{K. K\"onig}}
\altaffiliation[Present address: ]{National Superconducting Cyclotron Laboratory, Michigan State University, East Lansing, Michigan 48824, USA}
\affiliation{Institut f\"ur Kernphysik, Technische Universit\"at Darmstadt, D-64289 Darmstadt, Germany}

\author{S. Malbrunot-Ettenauer}
\affiliation{Experimental Physics Department, CERN, CH-1211 Geneva 23, Switzerland}
\author{T. Miyagi}
\affiliation{TRIUMF, 4004 Wesbrook Mall, Vancouver, British Columbia, V6T 2A3, Canada}
\author{R. Neugart}
\affiliation{Max-Planck-Institut f\"ur Kernphysik, D-69117 Heidelberg, Germany}
\affiliation{Institut f\"ur Kernchemie, Universit\"at Mainz, D-55128 Mainz, Germany}
\author{G. Neyens}
\affiliation{Experimental Physics Department, CERN, CH-1211 Geneva 23, Switzerland}
\affiliation{KU Leuven, Instituut voor Kern- en Stralingsfysica, 3001 Leuven, Belgium}
\author{W. N\"ortersh\"auser}
\affiliation{Institut f\"ur Kernphysik, Technische Universit\"at Darmstadt, D-64289 Darmstadt, Germany}
\author{\mbox{R. S\'anchez}}
\affiliation{GSI Helmholtzzentrum f\"ur Schwerionenforschung, D-64291 Darmstadt, Germany}
\author{F. Sommer}
\affiliation{Institut f\"ur Kernphysik, Technische Universit\"at Darmstadt, D-64289 Darmstadt, Germany}
\author{L.V. Rodr\'iguez}
\affiliation{Max-Planck-Institut f\"ur Kernphysik, D-69117 Heidelberg, Germany}
\affiliation{Institut de Physique Nucl\'eaire, CNRS-IN2P3, Universit\'e Paris-Sud, Universit\'e Paris-Saclay, 91406 Orsay, France}
\author{L. Xie}
\affiliation{School of Physics and Astronomy, The University of Manchester, Manchester, M13 9PL, United Kingdom}
\author{Z.Y. Xu}
\affiliation{KU Leuven, Instituut voor Kern- en Stralingsfysica, 3001 Leuven, Belgium}
\author{X.F. Yang}
\affiliation{KU Leuven, Instituut voor Kern- en Stralingsfysica, 3001 Leuven, Belgium}
\affiliation{School of Physics and State Key Laboratory of Nuclear Physics and Technology, Peking University, Beijing 100871, China}
\author{D.T. Yordanov}
\affiliation{Institut de Physique Nucl\'eaire, CNRS-IN2P3, Universit\'e Paris-Sud, Universit\'e Paris-Saclay, 91406 Orsay, France}

\date{\today}

\begin{abstract}
Hyperfine spectra of $^\text{27-32}$Al ($Z=13$) have been measured at the ISOLDE-CERN facility via collinear laser spectroscopy using the $3s^23p\ ^2\text{P}^\text{o} _{3/2}\rightarrow 3s^24s\ ^2\text{S}_{1/2}$ atomic transition. For the first time, mean-square charge radii of radioactive aluminum isotopes have been determined alongside the previously unknown magnetic dipole moment of $^{29}$Al and electric quadrupole moments of $^{29,30}$Al. A potentially reduced charge radius at $N=19$ may suggest an effect of the $N=20$ shell closure, which is visible in the Al chain, contrary to other isotopic chains in the $sd$ shell. 
\\ The experimental results are compared to theoretical calculations in the framework of the valence-space in-medium similarity renormalization group using multiple sets of two and three-nucleon forces from chiral effective field theory.
While the trend of experimental magnetic dipole and electric quadrupole moments is well reproduced, the absolute values are underestimated by theory, consistent with earlier studies. Moreover, both the scale and trend of the charge radii appear to be very sensitive to the chosen interaction.

\end{abstract}

\maketitle

\setlength{\abovedisplayskip}{5pt}
\setlength{\belowdisplayskip}{8pt}
\setlength{\abovedisplayshortskip}{0pt}
\setlength{\belowdisplayshortskip}{0pt}

	\begin{figure*}[t]
	\includegraphics[width=0.75\textwidth]{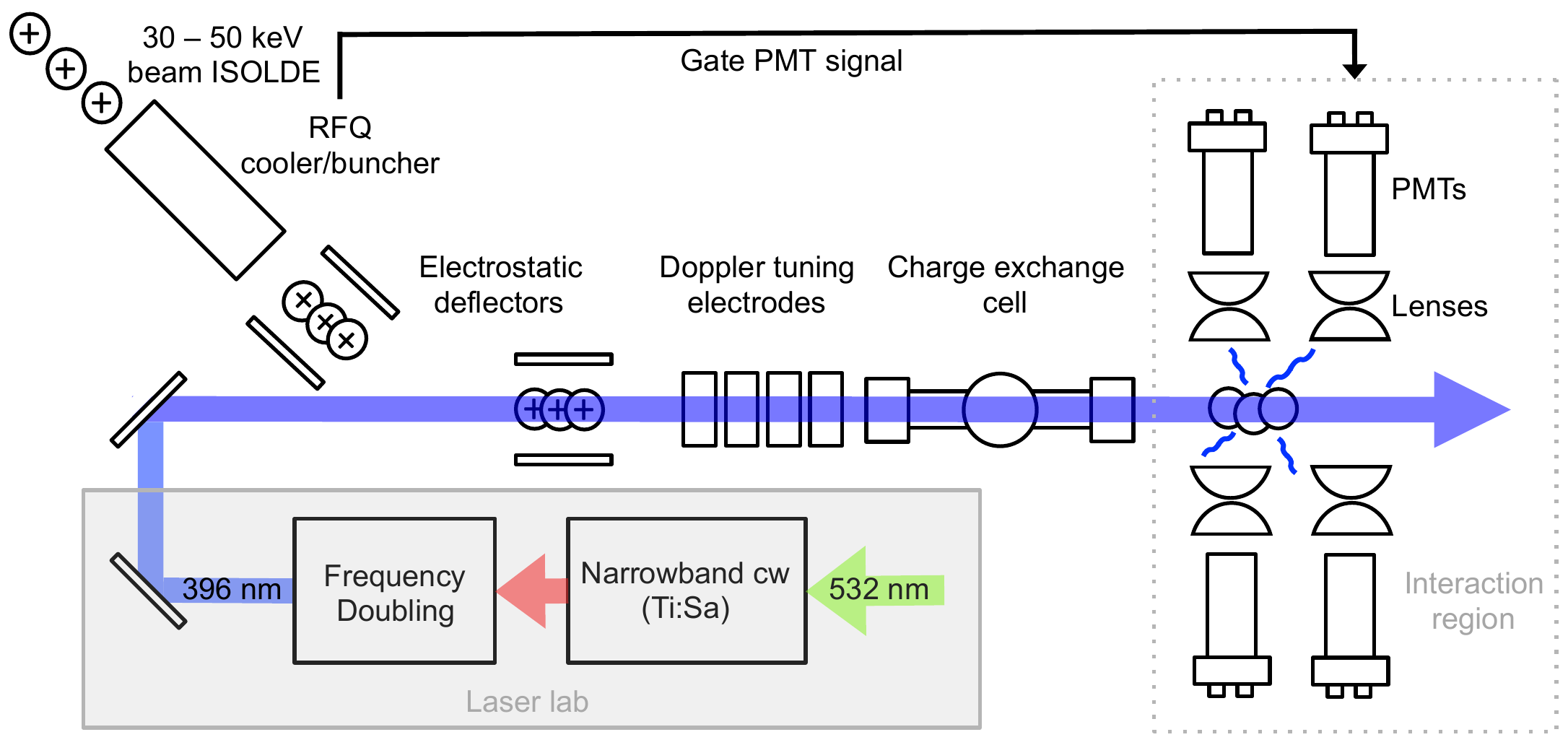}
	\caption{Schematic depiction of the COLLAPS collinear laser spectroscopy setup at ISOLDE-CERN. More details can be found in the text.} 
	\label{Fig:COLLAPS}
	\end{figure*}

\section{Introduction}
Laser spectroscopy performed at radioactive ion beam facilities is a powerful technique to study isotopes all across the chart of nuclei \cite{Campbell2015}. By measuring atomic (or ionic) hyperfine spectra, ground and isomeric state spins, electromagnetic moments and differences in mean-square charge radii can be obtained in a nuclear-model independent way. Since these observables each probe different nuclear structure aspects, laser spectroscopic studies are ideal to systematically investigate the performance of state-of-the-art nuclear theories.
\\ In the last few decades, tremendous progress has been made in solving the nuclear many-body problem from first principles. Chiral effective field theory~\cite{Epel09RMP,Mach11PR}
provides microscopic interactions rooted in QCD, the fundamental theory of the strong interaction, while novel methods to solve the many-body Schr\"odinger equation in medium- and heavy-mass nuclei have been developed. Combined with the advances in computing power, a rapidly growing number of nuclides can be described within such an \textit{ab-initio} framework \cite{Hergert2016}. Particularly, the Valence-Space In-Medium Similarity Renormalization Group (\mbox{VS-IMSRG})  method \cite{Tsukiyama2012,Bogner2014SM,Stroberg2017,Stroberg2019} has emerged as a versatile tool to study open-shell nuclei in the medium-mass region. Its scope now reaches up to the tin region \cite{Morris2018,Stroberg2019,Manea2020} and recently also isotopes requiring multi-shell valence spaces, e.g.~nuclei in islands of inversion, have become accessible \cite{Miyagi2020}. Not limited to \mbox{(near-)magic} nuclei, the VS-IMSRG technique allows the exploration of the microscopic origins of global features such as the driplines~\cite{Holt2019}, new magic numbers~\cite{Taniuchi2019} and Gamow-Teller quenching~\cite{Gysb19GT}, or local features along an isotopic chain, e.g.\  odd-even staggering in binding energies and radii \cite{Degroote2020} or the development of $E2$-strength \cite{Henderson2018,Henderson2020}. 
These studies illustrate the importance of observables beyond binding energies to gain complementary insights in the validity of a theoretical approach. Nevertheless, a comprehensive investigation of a wide range of properties and isotopes using \textit{ab-initio} machinery has only just begun.
\\ In this article, we focus on the magnetic dipole moments, $\mu$, electric quadrupole moments, $Q_s$, and changes in mean-square charge radii, $\delta \langle r^2 \rangle$, of the ground states of the aluminum isotopes $^\text{27-32}$Al at $Z=13$. Because a proper treatment of correlations is important to describe the structure of mid-shell nuclei like Al, these isotopes
 are interesting candidates to gauge the performance of the latest advances in \mbox{VS-IMSRG} calculations.
\\ Experimentally, the aluminum isotopic chain has been studied extensively in the past (see e.g.~\cite{Minamisono1981,Morton2002,Himpe2006,Kameda2007,Kwiatkowski2015,Heylen2016,Xu2018}) but the magnetic moment of $^{29}$Al and quadrupole moments of $^{29,30}$Al near stability were not yet determined. Furthermore, the collinear laser spectroscopy experiment performed in this work provides for the first time charge radii of radioactive Al isotopes. In particular, the development of these radii towards the $N=20$ shell closure is of interest. 

	\begin{figure}[t!]
	\includegraphics[width=0.9\columnwidth]{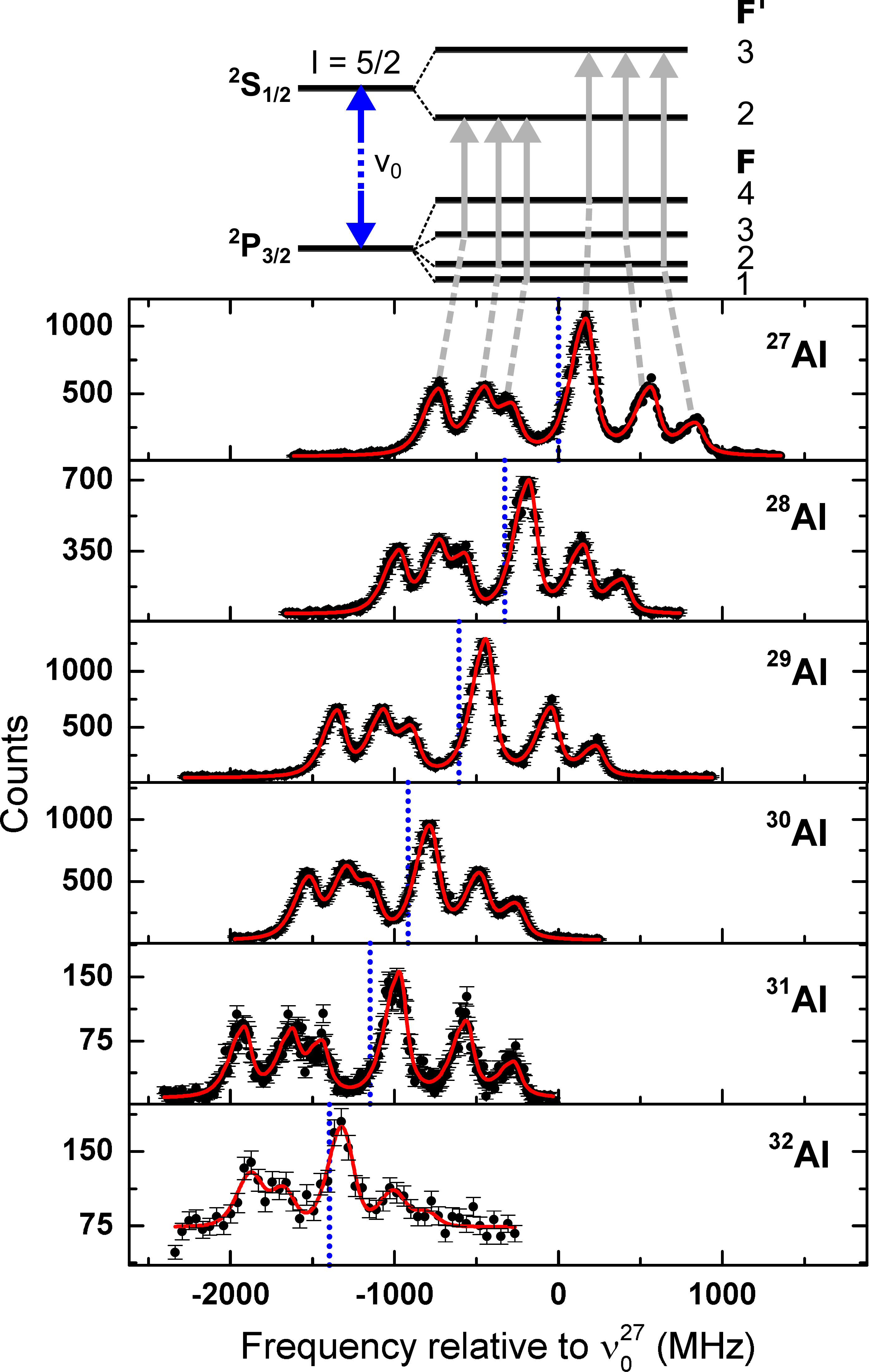}
	\caption{Fluorescence counts as a function of frequency obtained for $^\text{27-32}$Al on the $3s^23p\ ^2\text{P}^\text{o}_{3/2}\rightarrow 3s^24s\ ^2\text{S}_{1/2}$ atomic transition. The red line represents the best fit. The centroid frequency $\nu_0$ for each isotope is indicated with a vertical dashed line. On top of the figure, the hyperfine levels of an $I = 5/2$ nucleus like $^{27}$Al are shown and the transitions probed using laser spectroscopy are indicated with grey arrows.} 
	\label{Fig:HFS}
	\end{figure}

\section{Experimental details}

The experiment has been performed at the \mbox{COLLAPS} collinear laser spectroscopy beam line \cite{Neugart2017} at ISOLDE-CERN \cite{Borge2017}. A schematic of the set-up can be seen in Fig.~\ref{Fig:COLLAPS} and more details can be found in Refs.~\cite{Kreim2014,Neugart2017}. Neutral atoms of $^\text{27-32}$Al were produced by bombarding a uranium carbide target with 1.4-GeV protons from the PS booster. These isotopes were subsequently extracted from the target, ionised by ISOLDE's resonance ionization laser ion source (RILIS) \cite{Koster2003}, accelerated to 30 keV and mass separated. Afterwards, the ion beam passed through ISCOOL \cite{Franberg2008}, a He-buffer-gas-filled radiofrequency quadrupole (RFQ) cooler and buncher, where the ions were accumulated for approximately 50 ms and released in time bunches of a few $\mu$s. These ion bunches were then guided to the \mbox{COLLAPS} beam line where they were spatially overlapped with a co-propagating laser beam. Before neutralising the singly-charged ions to neutral atoms in an alkali-vapor-filled charge exchange cell, a Doppler-tuning voltage was applied to alter the ions' velocity and, hence, to scan the laser frequency in the reference frame of the atoms. Laser spectroscopy was performed on the $3s^23p\ ^2\text{P}^\text{o}_{3/2}\rightarrow 3s^24s\ ^2\text{S}_{1/2}$ atomic transition at \mbox{25 235.696 cm$^{-1}$}. This  transition was probed with around 2 mW of continuous-wave (cw) laser light generated by the frequency-doubled output of a titanium-sapphire ring laser. Laser-induced fluorescence was detected by a light collection system placed around the laser-atom interaction region. It consists of two rows of two photomultiplier tubes (PMT), each associated with their corresponding imaging lens arrangement \cite{Kreim2014}. By gating the fluorescence signals on the passage of the atom bunches through the interaction region, background from \mbox{continuously} scattered laser light and PMT dark counts was suppressed by four orders of magnitude.

\section{Results}

Experimental hyperfine spectra of $^\text{27-32}$Al are shown in Fig.~\ref{Fig:HFS}. As illustrated for $^{27}$Al in the first panel of the figure, each resonance corresponds to a transition between a hyperfine level of the $^2\text{P}^\text{o}_{3/2}$ lower and $^2\text{S}_{1/2}$ upper state level. The position of these resonances $\nu_{F,F'}$ is described by
\begin{align*}
\nu_{F,F'} = \nu_0 + \Delta \nu_{F'} - \Delta \nu_{F}
\end{align*} 
where $\nu_0$ is the unperturbed transition frequency between the fine structure levels, referred to as the centroid frequency. $\Delta \nu_{F}$ and $\Delta \nu_{F'}$ are the frequency differences between the hyperfine states and their respective fine structure levels. These depend on
the hyperfine parameters $A_J$ and $B_J$ for each atomic state $J$ according to
\begin{align*}
\Delta \nu_{F} = A_J\frac{C}{2} + B_J \frac{3C(C+1) - I(I+1)J(J+1)}{8(2I-1)(2J-1)IJ}.
\end{align*}
Here,  $I$, $J$ and $F$ ($\boldsymbol{F} = \boldsymbol{I}+\boldsymbol{J}$) are the nuclear, atomic and total angular momenta, respectively, and $C = F(F+1)-I(I+1)-J(J+1)$. \\ 
The SATLAS analysis library \cite{Gins2018} was used to fit the hyperfine structures using a $\chi^2$-minimisation procedure. In the fit, the centroid $\nu_0$, and hyperfine parameters $A(\text{P}^\text{o}_{3/2})$ and $B(\text{P}^\text{o}_{3/2})$ were taken as free parameters while the ratio $A(\text{S}_{1/2})/A(\text{P}^\text{o}_{3/2})= 4.5701$ was fixed to the value determined for the $^{27}$Al reference isotope (obtained with an uncertainty of 0.0014). The intensities of the individual hyperfine peaks were left to vary freely in the fit. The asymmetric peak profiles, related to inelastic processes in the charge exchange cell \cite{Bendali1986,Klose2012114, Kreim2014}, were found to be best described by a Voigt profile including four satellite peaks at an empirically determined energy offset. The relative intensities of these satellite peaks were constrained assuming Poisson's law.  
\\ For $^{27}$Al, a statistically reasonable agreement with the known hyperfine parameters of the $\text{P}^\text{o}_{3/2}$ state was obtained \cite{Lew1949,Weber1987,Levins1997}, see Table \ref{Table:HFparam-moments}. The precision on the $A(\text{S}_{1/2}) = 431.11(9)$ MHz value of $^{27}$Al deduced in this work was improved by more than two orders of magnitude as compared to the low-precision value in Ref.~\cite{Jiang1982}.

\subsection{Nuclear moments}
\begin{table*}[t!] 
\caption{Measured hyperfine $A(\text{P}^\text{o}_{3/2})$ and $B(\text{P}^\text{o}_{3/2})$ parameters and corresponding magnetic dipole moments $\mu$ and electric quadrupole moments $Q_s$ for $^\text{27-32}$Al ground states. The hyperfine anomaly is expected to be small relative to the experimental precision and neglected in the extraction of $\mu$. The reference moments from $^{27}$Al are taken from Refs.~\cite{Antusek2015, Kello1999}, which take into account recent atomic and molecular calculations of the hyperfine magnetic field and electric field gradients, necessary to extract $\mu$ and $Q_s$, respectively. Literature values for the other isotopes are re-evaluated with respect to these reference values in Refs.~\cite{Stone2019,Derydt2013}. References to the original experimental papers can be found in there.}
\begin{ruledtabular}
\begin{tabular}{c c c c l l l l l l }
\\ [-2.3ex]
$A$ & $N$ & $I^{\pi}$ & $T_{1/2}$ & $A(\text{P}^\text{o}_{3/2})$ (MHz) & $B(\text{P}^\text{o}_{3/2})$ (MHz) & $\mu_\text{exp}$ $(\mu_\text{N})$ & $\mu_\text{lit}$ $(\mu_\text{N})$ & $Q_{s,\text{exp}}$ (fm$^2$) & $Q_{s,\text{lit}}$ (fm$^2$)  \\
 [0.45ex]
\colrule \\ [-1.85ex]
27 & 14 &5/2$^+$ & stable & +94.33\,(4)\footnotemark[1] & +18.1\,(2)\footnotemark[1] & \multicolumn{1}{l}{\emph{Reference}} & +3.64070\,(2) &\multicolumn{1}{l}{\emph{Reference}} &  +14.66(10) \\
28 & 15 &3$^+$ & 2.24 m & +70.07\,(6) & +18.1\,(8) & +3.245\,(3)&   \hspace{0.25cm}3.241\,(5)& +14.7\,(7) &  \hspace{0.25cm}17.2\,(12)  \\
29 & 16 &5/2$^+$ & 6.56 m & +94.97\,(5) & +18.2\,(6) & +3.665\,(2)&   \multicolumn{1}{c}{-}& +14.8\,(5) & \multicolumn{1}{c}{-}  \\
30 & 17 &3$^+$ & 3.62 s & +65.36\,(7) & +14.9\,(10) & +3.027\,(4)&  \hspace{0.25cm}3.012\,(7) & +12.1\,(8) & \multicolumn{1}{c}{-}  \\
31 & 18 &5/2$^+$ & 644 ms & +99.0\,(3) & +19.3\,(17) & +3.822\,(11)&  \hspace{0.25cm}3.832\,(5) & +15.6\,(14) & \hspace{0.25cm}13.40\,(16)  \\
32 & 19 &1$^+$ & 33 ms & +124\,(3) & +\hspace{0.15cm}2\,(6) & +1.92\,(4) &  \hspace{0.25cm}1.953\,(2)& +\hspace{0.15cm}1(5) & \hspace{0.40cm}2.5\,(2)  \\
\end{tabular}
\end{ruledtabular}
\footnotetext[1]{To be compared with the average of the literature values: $A(\text{P}^\text{o}_{3/2})=94.25(4)$ MHz  and $B(\text{P}^\text{o}_{3/2})=18.8(3)$ MHz \cite{Lew1949,Weber1987,Levins1997}.}
\label{Table:HFparam-moments}
\end{table*}

Table \ref{Table:HFparam-moments} shows the measured hyperfine $A(\text{P}^\text{o}_{3/2})$ and $B(\text{P}^\text{o}_{3/2})$ parameters of $^\text{27-32}$Al along with the magnetic dipole moments, $\mu$, and electric quadrupole moments, $Q_s$, extracted according to 
\begin{equation*}
\mu= \mu_\text{ref}\frac{AI}{A_\text{ref}I_\text{ref}}, \ Q_s = Q_{s,\text{ref}}\frac{B}{B_\text{ref}}.
\end{equation*}
Reference values for $^{27}$Al were taken from Refs.~\cite{Antusek2015, Kello1999} for the magnetic and quadrupole moment, respectively. The hyperfine anomaly was assumed to be negligible in the extraction of the magnetic moments. 
The present data provides for the first time an internally consistent set of Al moments determined with respect to a single reference isotope and measured in the same experimental conditions. This avoids potential discrepancies related to the applied shielding corrections or inconsistent electric field gradient calculations present when extracting magnetic and quadrupole moments from \mbox{($\beta$)-NMR} experiments \cite{Stone2019,Derydt2013}. No systematic deviations between our moments and the available literature values were observed.
\\ As seen from Table \ref{Table:HFparam-moments}, the previously unknown quadrupole moment of $^{30}$Al ($I = 3$) and magnetic and quadrupole moment of $^{29}$Al ($I = 5/2$) are similar to the moments of the Al isotopes with the same spin, suggesting a comparable nuclear structure. In general, it has been found that the aluminum ground states between $N = 14$ and $N = 19$ are well described within an $sd$-picture \cite{Himpe2006}.

\subsection{Isotope shifts and mean-square charge radii}\label{IS}
\begin{table}[t]  
\caption{Aluminum isotope shifts $\delta \nu^{27,A}$ measured in the $3s^23p\ ^2\text{P}^\text{o}_{3/2}\rightarrow 3s^24s\ ^2\text{S}_{1/2}$ transition. The relative mean-square charge radii $\delta \langle r^2 \rangle^{27,A}$ with respect to $^{27}$Al are extracted from these isotope shifts using $M = -243$ GHz u and \mbox{$F = 76.2$ MHz/fm$^2$}. Systematic uncertainties due to a $1.5 \cdot 10^{-4}$ relative uncertainty on the beam energy are indicated with square brackets, while the uncertainties on the $\delta \langle r^2 \rangle^{27,A}$ arising from the atomic calculations of $M$ and $F$ are shown in curly parentheses.}
\begin{ruledtabular}
\begin{tabular}{c c l l}
\\ [-2.3ex]
$A$ & $N$ & $\delta \nu^{27,A}$ (MHz) & $\delta \langle r^2 \rangle^{27,A}$ (fm$^2$) \\
 [0.45ex]
\colrule \\ [-1.85ex]
27 & 14 & \hspace{0.72cm}0 & \hspace{0.72cm} 0 \\
28 & 15  & $-$\hspace{0.15cm}321.9\,(8)[33] &+ 0.003\,(10)[43]\{72\} \\
29 & 16 & $-$\hspace{0.15cm}610.7\,(6)[64] & +0.141\,(8)[84]\{134\}\\
30 & 17 & $-$\hspace{0.15cm}889.5\,(12)[101] & +0.164\,(15)[132]\{196\}\\
31 & 18 & $-$1141.0\,(12)[136] &+0.301\,(16)[178]\{250\} \\
32 & 19 & $-$1401\,(7)[17]& +0.12\,(9)[22]\{31\}\\
\end{tabular}
\end{ruledtabular}
\label{Table:IS}
\end{table}

The measured isotope shifts, defined as \mbox{$\delta \nu^{27,A} = \nu_0^A- \nu_0^{27}$}, are shown in Table \ref{Table:IS}. By alternating measurements of radioactive isotopes with those of the stable $^{27}$Al reference, effects from drifts in experimental conditions largely cancel out. A systematic uncertainty on the deduced isotope shifts accounts for a $1.5\cdot 10^{-4}$ relative uncertainty on the beam energy.
\\ From the measured isotope shifts, differences in mean-square charge radii, $\delta\langle r^2\rangle^{27,A} = \langle r^2\rangle^{A} - \langle r^2\rangle^{27}$, can be extracted via \cite{MartenssonPendrill1990}
\begin{align*}
\delta \nu^{27,A}  = F\delta\langle r^2\rangle^{27,A} +M\frac{m_A - m_{27}}{m_{27}(m_A+m_e)},
\end{align*}
where $m_A$ and $m_{27}$ are the nuclear masses obtained by subtracting 13 electron masses from the atomic masses and $m_e$ is the electron mass. The atomic field shift and mass shift factors $F$ and $M$, respectively, can be empirically calibrated via a King-plot procedure if the absolute charge radii of at least three isotopes are known from other techniques, see for example Ref.~\cite{Cheal2012}. For chemical elements with a single stable isotope such as Al, only one absolute radius is known and one has to rely on atomic calculations instead. In these cases, a precise extraction of $\delta \langle r^2 \rangle$ is often challenging since the mass shift, which dominates the isotope shift for relatively light elements like Al \cite{Noertershaeuser2014}, is difficult to evaluate.\\
	
An in-depth investigation of the atomic factors for the $3s^23p\ ^2\text{P}^\text{o}_{3/2}\rightarrow 3s^24s\ ^2\text{S}_{1/2}$ transition has previously been performed in the multiconfiguration Dirac-Hartree-Fock framework \cite{Filippin2016}. Two computational techniques (RATIP and RIS3) were adopted to study  different electron correlation models: the core-valence + valence-valence (CV+VV) and core-valence + valence-valence + core-core correlations (CV+VV+CC), respectively. After analysis of the normal and specific contributions to the mass shift factor, it was found that the CV+VV model was the most reliable. Additionally, the effect of a common or separate optimisation of the orbital basis sets for the lower and upper atomic states was explored. Although the CC effects were more balanced in the common optimisation strategy, a separate optimisation was necessary to properly treat orbital relaxation (i.e.~to allow for an independent reorganisation of the spectator electrons of the lower and upper states during the excitation process). Within the CV+VV correlation model, the separate optimisation strategy was therefore preferred. Based on these considerations, in this work we selected the results of the CV+VV correlation model with separate optimisation of basis states. The spread between the RIS3 ($F=74$ MHz/fm$^2$ and $M = -239$ GHz u) and RATIP ($F=78.4$ MHz/fm$^2$ and $M = -247$ GHz u) computational methods is a good indication of the uncertainty on the calculations. Note that due to a different sign convention, the sign of $M$ is opposite here as compared to Ref.~\cite{Filippin2016}. 
 \begin{figure}
  \includegraphics[width=\columnwidth]{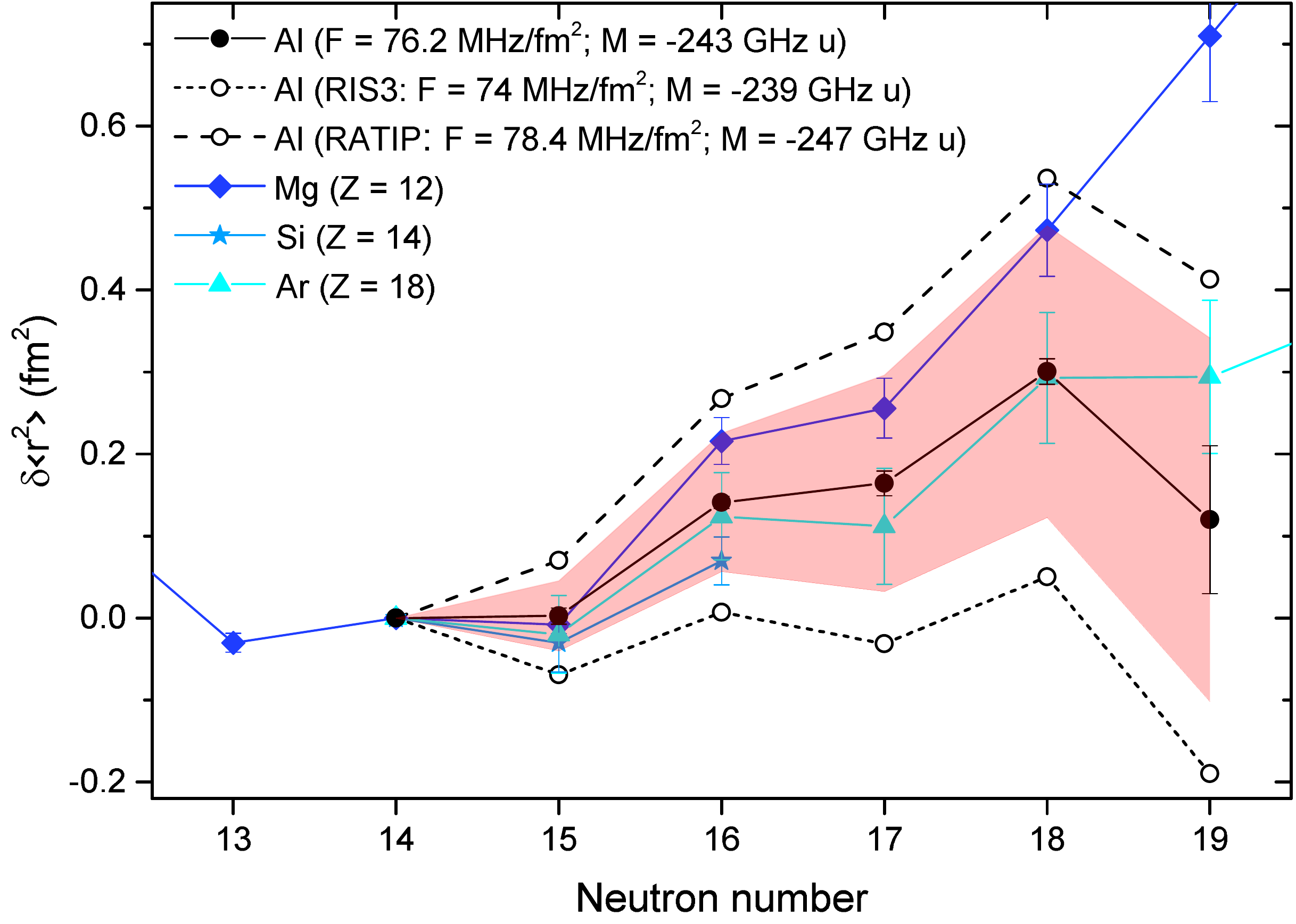}
  \caption{Changes in mean-square charge radii of Al isotopes with respect to $N=14$ obtained by combining the isotope shifts measured in this work with the atomic field and mass shift factors $F = 76.2$ MHz/fm$^2$ and $M = -243$ GHz u determined in Ref.~\cite{Filippin2016}, see text for details. The red shaded band indicates the systematic uncertainty due to a $1.5 \cdot 10^{-4}$ relative uncertainty on the beam energy. Results using the $F$ and $M$ for the RIS3 and RATIP computational methods are separately indicated by the dotted and dashed lines, respectively. Data for neighbouring isotopic chains are taken from \cite{Yordanov2012,Angeli2013,Blaum2008}. The charge radii of Na ($Z=11$) are not included in the plot because of the large systematic uncertainty on the slope \cite{Fricke2004}.}
  \label{Fig:chargeradii-FM}
    \end{figure}
\\ Figure \ref{Fig:chargeradii-FM} shows the changes in mean-square charge radii of Al obtained this way alongside the experimental charge radii of chemical elements below and above Al. The good agreement between the Al radii and the regional systematics supports that the relevant physics is captured in the atomic calculations. For the extraction of the final $\delta \langle r^2 \rangle$, see Table \ref{Table:IS}, the average between the two computational methods was adopted: $M = -243 \pm 4$ GHz u and $F = 76.2 \pm 2.2$ MHz/fm$^2$. Here, $\pm$ refers to the range determined by the two methods, rather than to a $1\sigma$ uncertainty interval. This range established a systematic uncertainty on the extracted results. Additionally, the uncertainty on the isotope shift due to incomplete knowledge of the beam energy was propagated to the charge radii, shown by the shaded red band in Fig.~\ref{Fig:chargeradii-FM}. It is important to note that the systematic uncertainties, due to the beam energy as well as the atomic factors separately, act the same way and result in a slope of the charge radii which can change as a whole, but which does not affect the local details, like odd-even staggering and discontinuities.

	\begin{figure*}
	\includegraphics[width=0.8\textwidth]{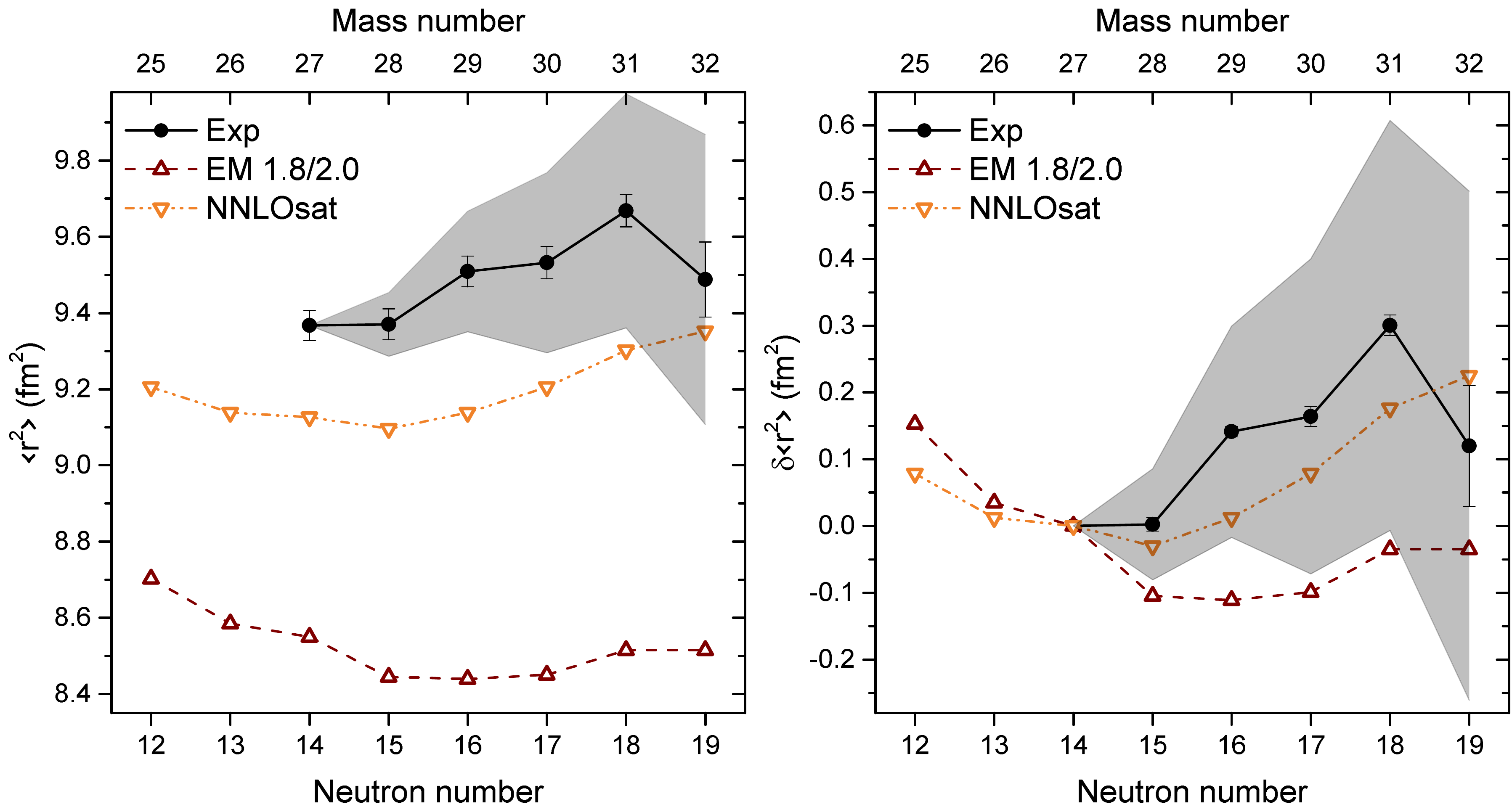}
	\caption{Absolute mean-square charge radii $\langle r^2 \rangle$ and differences in mean-square charge radii $\delta\langle r^2 \rangle$ along the aluminum isotopic chain. Experimental results are compared to VS-IMSRG calculations. The gray band indicates the uncertainty on the slope of the charge radii arising due to the uncertainties on the atomic factors as well as the beam energy, as explained in sec.~\ref{IS}.} 
	\label{Fig:radii}
	\end{figure*}
Absolute mean-square charge radii $\langle r^2 \rangle$ along the isotopic chain were extracted by combining our values for the differences in mean-square charge radii $\delta \langle r^2 \rangle$ with the absolute mean-square charge radius of $^{27}$Al. To obtain the latter, experimental data from muonic atom spectroscopy were combined with elastic electron scattering measurements according to the procedure laid out in Ref.~\cite{Fricke2004}. We started from the Barrett equivalent radius \mbox{$R^{\mu}_{k\alpha} = 3.9354(24)$ fm} deduced from the $2p-1s$ transition energy in the muonic atom \cite{Fricke2004}. In order to extract a model-independent root mean-square radius $\sqrt{\langle r^2 \rangle^{\mu e}}$, this Barrett equivalent radius was divided by the ratio of radial moments $V^e_2 = \displaystyle\frac{R^{e}_{k\alpha}}{\sqrt{\langle r^2 \rangle^e}}$ using the same values for $\alpha$ and $k$ as in the muonic data. In the evaluation of both $R^{e}_{k\alpha}$ and $\langle r^2 \rangle^e$, the charge density distribution, $\rho(r)$, measured in electron scattering experiments was used.  Specifically, $\rho(r)$ determined from the Fourier-Bessel coefficients measured by Rothhaas and collaborators  tabulated in Ref.~\cite{Devries1987} has been chosen since it has the benefit of being model-independent. Using this form of density distribution gave a value of $V^e_2 = 1.2858(26)$. An accurate evaluation of the error on this value would require full knowledge of the uncertainty matrix associated with the calculation of these coefficients, which was unfortunately not available. Instead, the uncertainty was conservatively estimated by calculating the $V^e_2$ values using model-dependent charge density distributions determined with data from two different electron scattering experiments \cite{Lombard1967,Fey1973} yielding $V^e_2 = 1.2847$ and $V^e_2 = 1.2832$, respectively. The maximal difference between the $V^e_2$ values computed from the three data sets gave the final error. Following this procedure, a root mean-square charge radius of $\sqrt{\langle r^2 \rangle}=3.061(6)$ fm was obtained for  $^{27}$Al, corresponding to $\langle r^2 \rangle=9.37(4)$ fm$^2$.

\section{$\boldsymbol{\delta \langle r^2 \rangle}$ development towards $\boldsymbol{N = 20}$}

The mean-square charge radii relative to $^{27}$Al are presented in Fig.~\ref{Fig:radii}. These charge radii show a normal odd-even staggering on top of a generally increasing trend between $N=14$ and $N=18$. At $N=19$ however, the observed decrease in charge radius appears larger than expected from the odd-even staggering alone. Due to the relatively large uncertainty on the $^{32}$Al value, the deviation from a normally increasing trend is only around 2$\sigma$. Nevertheless, it is interesting to discuss briefly what such a decrease might implicate if confirmed by a more precise measurement. Typically, a local dip in the trend of charge radii is seen in the vicinity of magic numbers, which is related to the reduced correlations at closed shells. It would therefore be natural to interpret the decline at $N=19$ as the start of the dip leading up to the $N=20$ shell closure. However, $N=20$ is a peculiar case; such an effect is absent in the isotopic chains for which the charge radii are known so far. Above Al, the smoothly increasing charge radii of Ar, K and Ca ($Z = 18 - 20$) across $N=20$ have been interpreted as due to a balance between the monopole and the quadrupole proton-core polarisation effects when neutrons fill the $sd$ shell below $N = 20$ and the $f_{7/2}$ orbital above \cite{Blaum2008,Rossi2015}. Below Al on the other hand, a sudden increase in charge radii seen for Na and Mg ($Z=11,12$) is explained by the onset of deformation in the island of inversion around $N=20$ \cite{Huber1978,Touchard1982,Yordanov2012}. In this region, deformed intruder states in which neutrons are excited across $N=20$ into the $pf$-shell become the ground state below the normally expected (spherical) $sd$-states. The Al isotopes form the northern border of this island of inversion, with $^{33}$Al at $N=20$ being the transition point for which intruder configurations become significant \cite{Himpe2006,Heylen2016}. Based on these observations, a dip at $N=20$ in the Al charge radii trend would be quite unexpected. Hence, more precise measurements of the mean-square charge radii of $^{32}$Al and beyond are needed to clarify this issue.

	\begin{figure*}
	\includegraphics[width=0.9\textwidth]{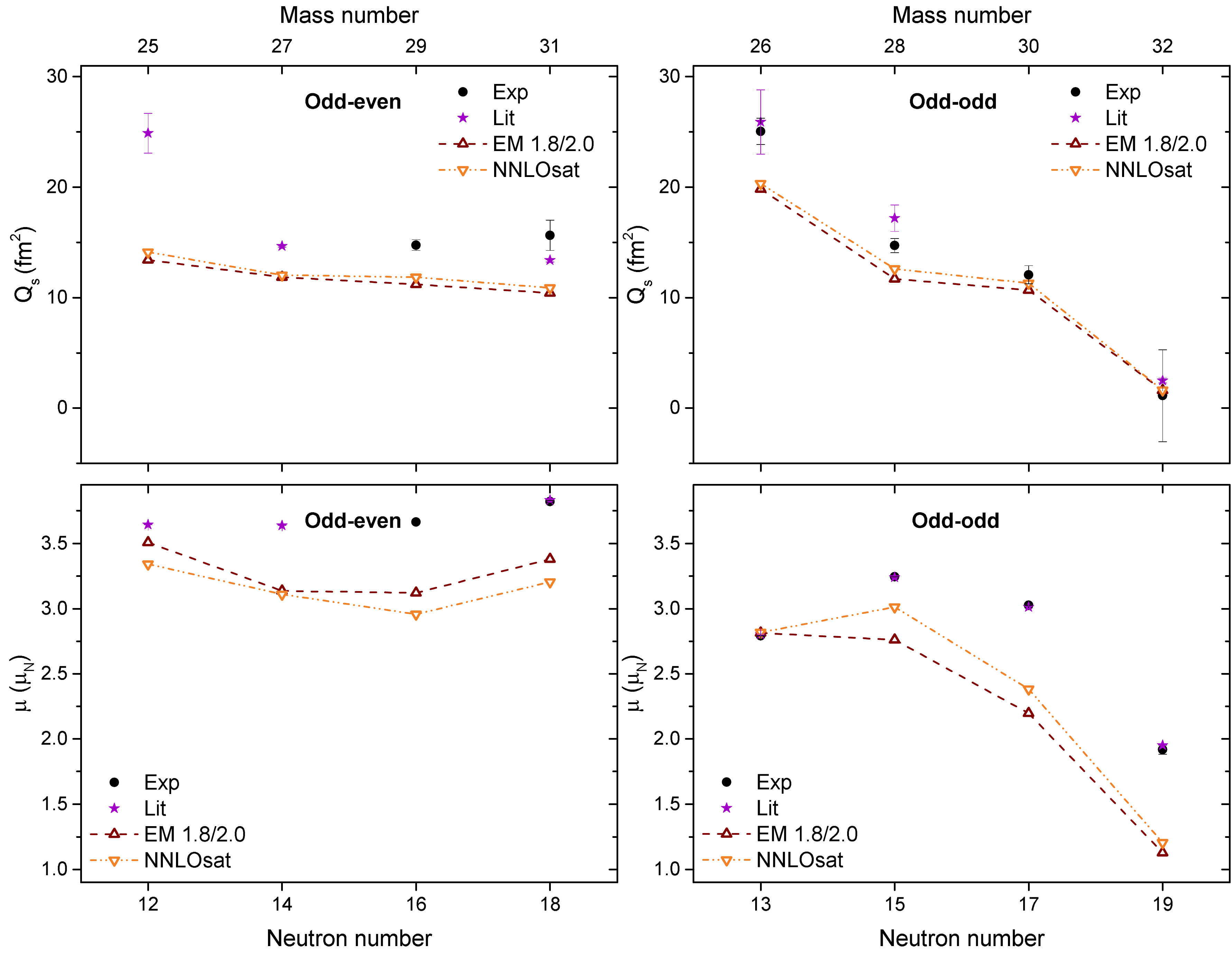}
	\caption{Electric quadrupole moments (top) and magnetic dipole moments (bottom) of $^\text{27-32}$Al. Experimental values obtained in this work and found in literature \cite{Stone2019,Derydt2013} are compared to calculations in the VS-IMSRG framework using the EM 1.8/2.0 and NNLO$_\text{sat}$ chiral interactions.} 
	\label{Fig:moments}
	\end{figure*}

\section{Comparison with VS-IMSRG results}	
We now compare the new measurements to calculations from the \textit{ab-initio} valence space in-medium similarity renormalization group method (VS-IMSRG) \cite{Tsukiyama2012,Bogner2014SM,Stroberg2019}. Since this method combines the broad applicability of the standard shell-model approach with microscopic interactions derived from two- and three-nucleon forces, it is an attractive option to study the structure of virtually all medium-mass isotopes, including open-shell isotopic chains like aluminum, in an \textit{ab-initio} framework.

\subsection{Details of the calculations}	
In the VS-IMSRG a computationally tractable valence-space Hamiltonian is decoupled from the much larger Hilbert space via an approximately unitary transformation~\cite{Hergert2016,Morris2015}, and the resulting effective Hamiltonian is diagonalized using traditional shell-model codes. As outlined in Ref.~\cite{Stroberg2017}, since the ensemble normal ordering procedure captures 3N forces between valence nucleons, a specific Hamiltonian is produced for each isotope separately. Other operators, including those for charge radii and electromagnetic moments, can be treated consistently in the same framework \cite{Simonis2017,Parzuchowski2017}. Note that the $M1$ operator currently used does not include contributions from meson-exchange currents. In principle, all many-body physics is captured in the decoupling procedure since all the excitations outside of the valence space are renormalized into the valence-space Hamiltonian and consistent operators. In practice, however, truncation of all operators in the IMSRG expansion at the two-body level, the IMSRG(2) approximation, is necessary to keep the problem computationally manageable. This means that higher-order terms, induced when deriving the effective operator, are not taken into account, which introduces some level of error in the procedure. Note that unlike in typical shell model calculations with empirically derived effective interactions, here bare charges and $g$-factors are used.

In this work, the IMSRG calculations are performed in a harmonic oscillator basis with \mbox{$\hbar\omega=16$ MeV} and quantum numbers $e=2n+l \leq e_\text{max}=12$. A further cut of $e_1+e_2+e_3\leq E_{3\text{max}} = 16$ is applied for the $3N$ matrix elements. Using the new multi-shell variation of the VS-IMSRG~\cite{Miyagi2020}, we are able to take $^{16}$O as a core and the valence space that includes the $sd$ proton orbitals and the $sdf_{7/2}p_{3/2}$ neutron orbitals. We also add the center-of-mass Hamiltonian to the initial Hamiltonian with the multiplier $\beta=3$, as detailed in Ref.~\cite{Miyagi2020} to separate the center-of-mass motion. The final valence-space-diagonalizations are performed with the KSHELL shell-model code \cite{Shimizu2019} and the effective operators were constructed using the imsrg++ code \cite{imsrgplusplus}.\\

We use two sets of NN+3N interactions derived from chiral effective field theory, \mbox{EM 1.8/2.0} \cite{Hebeler2011,Simonis2016,Simonis2017} and NNLO$_\text{sat}$ \cite{Eckstrom2015}. The \mbox{EM 1.8/2.0} interaction is only constrained by data of few-body systems ($A = 2,3,4$) and well reproduces binding energies to the $A=100$ region~\cite{Simonis2017,Morris2018}, including proton and neutron driplines~\cite{Holt2019}. However, it is known to generate charge radii which are consistently too small \cite{Simonis2017}. On the other hand, the NNLO$_\text{sat}$ interaction was optimised including a selected set of binding energies and radii of carbon and oxygen isotopes ($A\leq25$), on top of standard few-body data. This has improved the simultaneous reproduction of charge radii and binding energies, also for isotopes much heavier than those used in the optimisation \cite{GarciaRuiz2016,Lapoux2016,Hagen2015,Kaufmann2020}. 

\subsection{Nuclear moments and charge radii}
The calculated mean-square charge radii and nuclear moments together with our experimental results are shown in Figs.~\ref{Fig:radii} and \ref{Fig:moments}. In general, the observed trends are fairly well reproduced while the absolute scale deviates to varying extent as discussed next. Note that for the odd-odd isotopes, the state with the correct spin was not always calculated as the ground state but rather as an excited state below 850 keV. Although theoretical error bars are not yet available, this result is consistent with the 647 keV rms deviation found for VS-IMSRG calculations using the EM 1.8/2.0 interaction in the $sd$-shell \cite{Stroberg2019}. In the following discussion and corresponding plots, only states with the correct spins are considered.
\\ First, the absolute and differences in mean-square charge radii are shown in Fig.~\ref{Fig:radii}. Consistent with earlier findings \cite{Simonis2017,Degroote2020, Miyagi2020}, the absolute radii are very sensitive to the employed input interaction. As expected, the charge radii are underpredicted using the EM 1.8/2.0 interaction while the NNLO$_\text{sat}$ interaction generates radii in better, but not perfect, agreement with experiment. Moreover, the trend in $\delta \langle r^2 \rangle$ is different for both interactions. While the $\delta \langle r^2 \rangle$ obtained with EM 1.8/2.0 stay rather flat and do not reproduce the experimentally observed increasing trend, the slope of the NNLO$_\text{sat}$ is consistent within the present uncertainty on the experimental slope. Nevertheless, in contrast to the experimental charge radii, neither interaction yields an appreciable odd-even staggering. Up to now, calculations of mean-square charge radii of open-shell nuclei with the VS-IMSRG method are limited to Cu ($Z=29$) and Mn ($Z=25$) \cite{Simonis2017,Degroote2020} near the $Z=28$ shell closure. In these cases, results using the EM 1.8/2.0 interaction (or other interactions of the same family like PWA 2.0/2.0) could well reproduce the observed experimental trends. It is therefore remarkable that the description of the Al isotopes seem to be more challenging. A systematic investigation of charge radii in the VS-IMSRG method should help to better understand this issue in the future.
\\ Next, the quadrupole moments shown in the top row of Fig.~\ref{Fig:moments} are examined. With the exception of $^{25}$Al at $N=12$\footnote{Note that a verification of the quadrupole moment of $^{25}$Al, not remeasured in this work, would be advised due to the questionable quality of the obtained resonance in Ref.~\cite{Matsuta2007}.}, the quadrupole moments are underestimated by approximately 20\% across the isotopic chain and both the EM 1.8/2.0 and NNLO$_\text{sat}$ interactions give nearly identical results. This underprediction is fully in line with earlier studies of $E2$-observables including static quadrupole moments and  transition probabilities ($B(E2)$) \cite{Parzuchowski2017,Stroberg2019,Henderson2018,Henderson2020}, which identified the \mbox{IMSRG(2)} approximation as a major cause for the missing $E2$ strength. Due to the truncation of the operators at the two-body level, the effect of correlated multi-particle multi-hole pairs is underestimated. Furthermore, it was pointed out that the details of the input Hamiltonians do not have a large influence on the scale of the deviation. This conclusion is also supported here by the close similarity between the quadrupole moments obtained with both EM 1.8/2.0 and NNLO$_\text{sat}$ interactions. 
\\ Also for the magnetic moments in the bottom row of Fig.~\ref{Fig:moments}, both input interactions give comparable results which are in qualitative agreement with the experimental trend, while the absolute values are too small. So far, magnetic moments have only been studied intermittently in the VS-IMSRG approach \cite{Parzuchowski2017,Klose2019} and investigations into the origin of the observed deviation is still work in progress.
Since the meson-exchange currents show non-negligible effect on the magnetic moments in light nuclei~\cite{Pastore2013}, they will contribute at least partly to the discrepancy.
Furthermore, it is reasonable to assume that the IMSRG(2) approximation will also have an effect on the calculated magnetic moments, although different kinds of correlations might be important as compared to the quadrupole moment. Note that underestimated correlations could lead to magnetic moments which are either too small (like in this case) or too large (like e.g.~for $^{37,39}$Ca \cite{GarciaRuiz2015,Klose2019}). This is similar to the effect of introducing effective $g$-factors in phenomenological calculations.
Despite an overall deviation, the good reproduction of the trend for both the magnetic and quadrupole moments suggests that few-nucleon effects in the model are correctly taken into account.

\section{Summary}
The $^\text{27-32}$Al isotopes were studied via high resolution collinear laser spectroscopy at ISOLDE-CERN. State-of-the-art atomic physics calculations in combination with the isotope shifts measured in this work gave access to changes in mean-square charge radii of radioactive Al isotopes for the first time. An apparent reduction in the charge radius of $^{32}$Al was discussed in the context of a potential shell effect at $N=20$, although firm conclusions can not be made due to its relatively large uncertainty. Furthermore, our measurements of the magnetic dipole moment of $^{29}$Al and electric quadrupole moment of $^{29,30}$Al, fill the previously existing gap in nuclear moments near the valley of stability.
\\ Experimental magnetic moments, quadrupole moments and changes in mean-square charge radii of $^{25-32}$Al were compared to calculations within the VS-IMSRG approach using interactions derived from chiral effective field theory. No effective modifications to the $g$-factor and charge were introduced. Generally, the trends of the magnetic and quadrupole moments were well reproduced while absolute values were underestimated. The description of the mean-square charge radii proved to be more challenging, although calculations using the NNLO$_\text{sat}$ interaction are in agreement with the observed experimental slope. Because these observables are each sensitive to distinct features of the underlying nuclear structure, they are well-suited to provide complementary benchmarks for \textit{ab-initio} calculations.

\section*{Acknowledgements}

The authors would like to thank the ISOLDE technical teams for their support during the preparation and running of the experiment. We acknowledge the support from FWO-Vlaanderen, the Max-Planck Society, UK Science and Technology Facilities Council (STFC) grant ST/P004598/1, the BMBF under Contract Nos.~05P18RDCIA and 05P18RDFN, the Helmholtz International Center for FAIR (HIC for FAIR), the FWO and FNRS Excellence of Science Programme (EOS-O022818F) and STFC grant ST/P004423/1.
We would also like to thank J.~Simonis and P.~Navr\'atil for providing the 3N matrix element files and S.~R.~Stroberg for the imsrg++ code \cite{imsrgplusplus} used to perform these calculations. TRIUMF receives funding via a contribution through the National Research Council of Canada. This  work was further supported by NSERC. Computations were performed with an allocation of computing resources on Cedar at WestGrid and Compute Canada, and on the Oak Cluster at TRIUMF managed by the University of British Columbia department of Advanced Research Computing (ARC).


\bibliography{combinedreferences}

\begin{thebibliography}{70}%
\makeatletter
\providecommand \@ifxundefined [1]{%
 \@ifx{#1\undefined}
}%
\providecommand \@ifnum [1]{%
 \ifnum #1\expandafter \@firstoftwo
 \else \expandafter \@secondoftwo
 \fi
}%
\providecommand \@ifx [1]{%
 \ifx #1\expandafter \@firstoftwo
 \else \expandafter \@secondoftwo
 \fi
}%
\providecommand \natexlab [1]{#1}%
\providecommand \enquote  [1]{``#1''}%
\providecommand \bibnamefont  [1]{#1}%
\providecommand \bibfnamefont [1]{#1}%
\providecommand \citenamefont [1]{#1}%
\providecommand \href@noop [0]{\@secondoftwo}%
\providecommand \href [0]{\begingroup \@sanitize@url \@href}%
\providecommand \@href[1]{\@@startlink{#1}\@@href}%
\providecommand \@@href[1]{\endgroup#1\@@endlink}%
\providecommand \@sanitize@url [0]{\catcode `\\12\catcode `\$12\catcode
  `\&12\catcode `\#12\catcode `\^12\catcode `\_12\catcode `\%12\relax}%
\providecommand \@@startlink[1]{}%
\providecommand \@@endlink[0]{}%
\providecommand \url  [0]{\begingroup\@sanitize@url \@url }%
\providecommand \@url [1]{\endgroup\@href {#1}{\urlprefix }}%
\providecommand \urlprefix  [0]{URL }%
\providecommand \Eprint [0]{\href }%
\providecommand \doibase [0]{http://dx.doi.org/}%
\providecommand \selectlanguage [0]{\@gobble}%
\providecommand \bibinfo  [0]{\@secondoftwo}%
\providecommand \bibfield  [0]{\@secondoftwo}%
\providecommand \translation [1]{[#1]}%
\providecommand \BibitemOpen [0]{}%
\providecommand \bibitemStop [0]{}%
\providecommand \bibitemNoStop [0]{.\EOS\space}%
\providecommand \EOS [0]{\spacefactor3000\relax}%
\providecommand \BibitemShut  [1]{\csname bibitem#1\endcsname}%
\let\auto@bib@innerbib\@empty
\bibitem [{\citenamefont {Campbell}\ \emph {et~al.}(2015)\citenamefont
  {Campbell}, \citenamefont {Moore},\ and\ \citenamefont
  {Pearson}}]{Campbell2015}%
  \BibitemOpen
  \bibfield  {author} {\bibinfo {author} {\bibfnamefont {P.}~\bibnamefont
  {Campbell}}, \bibinfo {author} {\bibfnamefont {I.}~\bibnamefont {Moore}}, \
  and\ \bibinfo {author} {\bibfnamefont {M.R.}\ \bibnamefont {Pearson}},\
  }\bibfield  {title} {\enquote {\bibinfo {title} {Laser spectroscopy for
  nuclear structure physics},}\ }\href {\doibase 10.1016/j.ppnp.2015.09.003}
  {\bibfield  {journal} {\bibinfo  {journal} {Progress in Particle and Nuclear
  Physics}\ }\textbf {\bibinfo {volume} {86}},\ \bibinfo {pages} {127--180}
  (\bibinfo {year} {2015})}\BibitemShut {NoStop}%
\bibitem [{\citenamefont {Epelbaum}\ \emph {et~al.}(2009)\citenamefont
  {Epelbaum}, \citenamefont {Hammer},\ and\ \citenamefont
  {Mei{\ss}ner}}]{Epel09RMP}%
  \BibitemOpen
  \bibfield  {author} {\bibinfo {author} {\bibfnamefont {E.}~\bibnamefont
  {Epelbaum}}, \bibinfo {author} {\bibfnamefont {H.-W}\ \bibnamefont {Hammer}},
  \ and\ \bibinfo {author} {\bibfnamefont {U.-G.}\ \bibnamefont
  {Mei{\ss}ner}},\ }\bibfield  {title} {\enquote {\bibinfo {title} {{Modern
  Theory of Nuclear Forces}},}\ }\href {\doibase
  https://doi.org/10.1103/RevModPhys.81.1773} {\bibfield  {journal} {\bibinfo
  {journal} {Reviews of Modern Physics}\ }\textbf {\bibinfo {volume} {81}},\
  \bibinfo {pages} {1773} (\bibinfo {year} {2009})}\BibitemShut {NoStop}%
\bibitem [{\citenamefont {Machleidt}\ and\ \citenamefont
  {Entem}(2011)}]{Mach11PR}%
  \BibitemOpen
  \bibfield  {author} {\bibinfo {author} {\bibfnamefont {R.}~\bibnamefont
  {Machleidt}}\ and\ \bibinfo {author} {\bibfnamefont {D.~R.}\ \bibnamefont
  {Entem}},\ }\bibfield  {title} {\enquote {\bibinfo {title} {{Chiral effective
  field theory and nuclear forces}},}\ }\href {\doibase
  10.1016/j.physrep.2011.02.001} {\bibfield  {journal} {\bibinfo  {journal}
  {Physics Reports}\ }\textbf {\bibinfo {volume} {503}},\ \bibinfo {pages}
  {1--75} (\bibinfo {year} {2011})}\BibitemShut {NoStop}%
\bibitem [{\citenamefont {Hergert}\ \emph {et~al.}(2016)\citenamefont
  {Hergert}, \citenamefont {Bogner}, \citenamefont {Morris}, \citenamefont
  {Schwenk},\ and\ \citenamefont {Tsukiyama}}]{Hergert2016}%
  \BibitemOpen
  \bibfield  {author} {\bibinfo {author} {\bibfnamefont {H.}~\bibnamefont
  {Hergert}}, \bibinfo {author} {\bibfnamefont {S.K.}\ \bibnamefont {Bogner}},
  \bibinfo {author} {\bibfnamefont {T.D.}\ \bibnamefont {Morris}}, \bibinfo
  {author} {\bibfnamefont {A.}~\bibnamefont {Schwenk}}, \ and\ \bibinfo
  {author} {\bibfnamefont {K.}~\bibnamefont {Tsukiyama}},\ }\bibfield  {title}
  {\enquote {\bibinfo {title} {The in-medium similarity renormalization group:
  A novel ab initio method for nuclei},}\ }\href {\doibase
  https://doi.org/10.1016/j.physrep.2015.12.007} {\bibfield  {journal}
  {\bibinfo  {journal} {Physics Reports}\ }\textbf {\bibinfo {volume} {621}},\
  \bibinfo {pages} {165 -- 222} (\bibinfo {year} {2016})}\BibitemShut {NoStop}%
\bibitem [{\citenamefont {Tsukiyama}\ \emph {et~al.}(2012)\citenamefont
  {Tsukiyama}, \citenamefont {Bogner},\ and\ \citenamefont
  {Schwenk}}]{Tsukiyama2012}%
  \BibitemOpen
  \bibfield  {author} {\bibinfo {author} {\bibfnamefont {K.}~\bibnamefont
  {Tsukiyama}}, \bibinfo {author} {\bibfnamefont {S.K.}\ \bibnamefont
  {Bogner}}, \ and\ \bibinfo {author} {\bibfnamefont {A.}~\bibnamefont
  {Schwenk}},\ }\bibfield  {title} {\enquote {\bibinfo {title} {{In-Medium
  Similarity Renormalization Group for Open-Shell Nuclei}},}\ }\href {\doibase
  10.1103/PhysRevC.85.061304} {\bibfield  {journal} {\bibinfo  {journal}
  {Physical Review C}\ }\textbf {\bibinfo {volume} {85}},\ \bibinfo {pages}
  {061304} (\bibinfo {year} {2012})}\BibitemShut {NoStop}%
\bibitem [{\citenamefont {Bogner}\ \emph {et~al.}(2014)\citenamefont {Bogner},
  \citenamefont {Hergert}, \citenamefont {Holt}, \citenamefont {Schwenk},
  \citenamefont {Binder}, \citenamefont {Calci}, \citenamefont {Langhammer},\
  and\ \citenamefont {Roth}}]{Bogner2014SM}%
  \BibitemOpen
  \bibfield  {author} {\bibinfo {author} {\bibfnamefont {S.~K.}\ \bibnamefont
  {Bogner}}, \bibinfo {author} {\bibfnamefont {H.}~\bibnamefont {Hergert}},
  \bibinfo {author} {\bibfnamefont {J.~D.}\ \bibnamefont {Holt}}, \bibinfo
  {author} {\bibfnamefont {A.}~\bibnamefont {Schwenk}}, \bibinfo {author}
  {\bibfnamefont {S.}~\bibnamefont {Binder}}, \bibinfo {author} {\bibfnamefont
  {A.}~\bibnamefont {Calci}}, \bibinfo {author} {\bibfnamefont
  {J.}~\bibnamefont {Langhammer}}, \ and\ \bibinfo {author} {\bibfnamefont
  {R.}~\bibnamefont {Roth}},\ }\bibfield  {title} {\enquote {\bibinfo {title}
  {{Nonperturbative shell-model interactions from the in-medium similarity
  renormalization group}},}\ }\href {\doibase 10.1103/PhysRevLett.113.142501}
  {\bibfield  {journal} {\bibinfo  {journal} {Physical Review Letters}\
  }\textbf {\bibinfo {volume} {113}},\ \bibinfo {pages} {142501} (\bibinfo
  {year} {2014})},\ \Eprint {http://arxiv.org/abs/1402.1407} {arXiv:1402.1407}
  \BibitemShut {NoStop}%
\bibitem [{\citenamefont {Stroberg}\ \emph {et~al.}(2017)\citenamefont
  {Stroberg}, \citenamefont {Calci}, \citenamefont {Hergert}, \citenamefont
  {Holt}, \citenamefont {Bogner}, \citenamefont {Roth},\ and\ \citenamefont
  {Schwenk}}]{Stroberg2017}%
  \BibitemOpen
  \bibfield  {author} {\bibinfo {author} {\bibfnamefont {S.~R.}\ \bibnamefont
  {Stroberg}}, \bibinfo {author} {\bibfnamefont {A.}~\bibnamefont {Calci}},
  \bibinfo {author} {\bibfnamefont {H.}~\bibnamefont {Hergert}}, \bibinfo
  {author} {\bibfnamefont {J.~D.}\ \bibnamefont {Holt}}, \bibinfo {author}
  {\bibfnamefont {S.~K.}\ \bibnamefont {Bogner}}, \bibinfo {author}
  {\bibfnamefont {R.}~\bibnamefont {Roth}}, \ and\ \bibinfo {author}
  {\bibfnamefont {A.}~\bibnamefont {Schwenk}},\ }\bibfield  {title} {\enquote
  {\bibinfo {title} {Nucleus-dependent valence-space approach to nuclear
  structure},}\ }\href {\doibase 10.1103/PhysRevLett.118.032502} {\bibfield
  {journal} {\bibinfo  {journal} {Physical Review Letters.}\ }\textbf {\bibinfo
  {volume} {118}},\ \bibinfo {pages} {032502} (\bibinfo {year}
  {2017})}\BibitemShut {NoStop}%
\bibitem [{\citenamefont {Stroberg}\ \emph {et~al.}(2019)\citenamefont
  {Stroberg}, \citenamefont {Bogner}, \citenamefont {Hergert},\ and\
  \citenamefont {Holt}}]{Stroberg2019}%
  \BibitemOpen
  \bibfield  {author} {\bibinfo {author} {\bibfnamefont {S.}~\bibnamefont
  {Stroberg}}, \bibinfo {author} {\bibfnamefont {Scott}\ \bibnamefont
  {Bogner}}, \bibinfo {author} {\bibfnamefont {Heiko}\ \bibnamefont {Hergert}},
  \ and\ \bibinfo {author} {\bibfnamefont {J.~D.}\ \bibnamefont {Holt}},\
  }\bibfield  {title} {\enquote {\bibinfo {title} {Nonempirical interactions
  for the nuclear shell model: An update},}\ }\href {\doibase
  10.1146/annurev-nucl-101917-021120} {\bibfield  {journal} {\bibinfo
  {journal} {Annual Review of Nuclear and Particle Science}\ }\textbf {\bibinfo
  {volume} {69}},\ \bibinfo {pages} {307--362} (\bibinfo {year}
  {2019})}\BibitemShut {NoStop}%
\bibitem [{\citenamefont {Morris}\ \emph {et~al.}(2018)\citenamefont {Morris},
  \citenamefont {Simonis}, \citenamefont {Stroberg}, \citenamefont {Stumpf},
  \citenamefont {Hagen}, \citenamefont {Holt}, \citenamefont {Jansen},
  \citenamefont {Papenbrock}, \citenamefont {Roth},\ and\ \citenamefont
  {Schwenk}}]{Morris2018}%
  \BibitemOpen
  \bibfield  {author} {\bibinfo {author} {\bibfnamefont {T.~D.}\ \bibnamefont
  {Morris}}, \bibinfo {author} {\bibfnamefont {J.}~\bibnamefont {Simonis}},
  \bibinfo {author} {\bibfnamefont {S.~R.}\ \bibnamefont {Stroberg}}, \bibinfo
  {author} {\bibfnamefont {C.}~\bibnamefont {Stumpf}}, \bibinfo {author}
  {\bibfnamefont {G.}~\bibnamefont {Hagen}}, \bibinfo {author} {\bibfnamefont
  {J.~D.}\ \bibnamefont {Holt}}, \bibinfo {author} {\bibfnamefont {G.~R.}\
  \bibnamefont {Jansen}}, \bibinfo {author} {\bibfnamefont {T.}~\bibnamefont
  {Papenbrock}}, \bibinfo {author} {\bibfnamefont {R.}~\bibnamefont {Roth}}, \
  and\ \bibinfo {author} {\bibfnamefont {A.}~\bibnamefont {Schwenk}},\
  }\bibfield  {title} {\enquote {\bibinfo {title} {Structure of the lightest
  tin isotopes},}\ }\href {\doibase 10.1103/PhysRevLett.120.152503} {\bibfield
  {journal} {\bibinfo  {journal} {Physical Review Letters}\ }\textbf {\bibinfo
  {volume} {120}},\ \bibinfo {pages} {152503} (\bibinfo {year}
  {2018})}\BibitemShut {NoStop}%
\bibitem [{\citenamefont {Manea}\ \emph {et~al.}(2020)\citenamefont {Manea}
  \emph {et~al.}}]{Manea2020}%
  \BibitemOpen
  \bibfield  {author} {\bibinfo {author} {\bibfnamefont {V.}~\bibnamefont
  {Manea}} \emph {et~al.},\ }\bibfield  {title} {\enquote {\bibinfo {title}
  {First glimpse of the \textit{N} = 82 shell closure below \textit{Z} = 50
  from masses of neutron-rich cadmium isotopes and isomers},}\ }\href {\doibase
  10.1103/PhysRevLett.124.092502} {\bibfield  {journal} {\bibinfo  {journal}
  {Physical Review Letters}\ }\textbf {\bibinfo {volume} {124}},\ \bibinfo
  {pages} {092502} (\bibinfo {year} {2020})}\BibitemShut {NoStop}%
\bibitem [{\citenamefont {Miyagi}\ \emph {et~al.}(2020)\citenamefont {Miyagi},
  \citenamefont {Stroberg}, \citenamefont {Holt},\ and\ \citenamefont
  {Shimizu}}]{Miyagi2020}%
  \BibitemOpen
  \bibfield  {author} {\bibinfo {author} {\bibfnamefont {T.}~\bibnamefont
  {Miyagi}}, \bibinfo {author} {\bibfnamefont {S.R.}\ \bibnamefont {Stroberg}},
  \bibinfo {author} {\bibfnamefont {J.~D.}\ \bibnamefont {Holt}}, \ and\
  \bibinfo {author} {\bibfnamefont {N.}~\bibnamefont {Shimizu}},\ }\bibfield
  {title} {\enquote {\bibinfo {title} {{Ab initio multi-shell valence-space
  Hamiltonians and the island of inversion}},}\ }\href@noop {} {\  (\bibinfo
  {year} {2020})},\ \Eprint {http://arxiv.org/abs/2004.12969}
  {arXiv:2004.12969} \BibitemShut {NoStop}%
\bibitem [{\citenamefont {Holt}\ \emph {et~al.}(2019)\citenamefont {Holt},
  \citenamefont {Stroberg}, \citenamefont {Schwenk},\ and\ \citenamefont
  {Simonis}}]{Holt2019}%
  \BibitemOpen
  \bibfield  {author} {\bibinfo {author} {\bibfnamefont {J.~D.}\ \bibnamefont
  {Holt}}, \bibinfo {author} {\bibfnamefont {S.R.}\ \bibnamefont {Stroberg}},
  \bibinfo {author} {\bibfnamefont {A.}~\bibnamefont {Schwenk}}, \ and\
  \bibinfo {author} {\bibfnamefont {J.}~\bibnamefont {Simonis}},\ }\bibfield
  {title} {\enquote {\bibinfo {title} {{Ab initio limits of atomic nuclei}},}\
  }\href@noop {} {\  (\bibinfo {year} {2019})},\ \Eprint
  {http://arxiv.org/abs/1905.10475} {arXiv:1905.10475} \BibitemShut {NoStop}%
\bibitem [{\citenamefont {Taniuchi}\ \emph {et~al.}(2019)\citenamefont
  {Taniuchi} \emph {et~al.}}]{Taniuchi2019}%
  \BibitemOpen
  \bibfield  {author} {\bibinfo {author} {\bibfnamefont {R.}~\bibnamefont
  {Taniuchi}} \emph {et~al.},\ }\bibfield  {title} {\enquote {\bibinfo {title}
  {{$^{78}$Ni revealed as a doubly magic stronghold against nuclear
  deformation}},}\ }\href {\doibase 10.1038/s41586-019-1155-x} {\bibfield
  {journal} {\bibinfo  {journal} {Nature}\ }\textbf {\bibinfo {volume} {569}},\
  \bibinfo {pages} {53--58} (\bibinfo {year} {2019})}\BibitemShut {NoStop}%
\bibitem [{\citenamefont {Gysbers}\ \emph {et~al.}(2019)\citenamefont {Gysbers}
  \emph {et~al.}}]{Gysb19GT}%
  \BibitemOpen
  \bibfield  {author} {\bibinfo {author} {\bibfnamefont {P.}~\bibnamefont
  {Gysbers}} \emph {et~al.},\ }\bibfield  {title} {\enquote {\bibinfo {title}
  {{Discrepancy between experimental and theoretical $\beta$-decay rates
  resolved from first principles}},}\ }\href {\doibase
  10.1038/s41567-019-0450-7} {\bibfield  {journal} {\bibinfo  {journal} {Nature
  Phys.}\ }\textbf {\bibinfo {volume} {15}},\ \bibinfo {pages} {428--431}
  (\bibinfo {year} {2019})}\BibitemShut {NoStop}%
\bibitem [{\citenamefont {de~Groote}\ \emph {et~al.}(2020)\citenamefont
  {de~Groote} \emph {et~al.}}]{Degroote2020}%
  \BibitemOpen
  \bibfield  {author} {\bibinfo {author} {\bibfnamefont {Ruben}\ \bibnamefont
  {de~Groote}} \emph {et~al.},\ }\bibfield  {title} {\enquote {\bibinfo {title}
  {Measurement and microscopic description of odd-even staggering of charge
  radii of exotic copper isotopes},}\ }\href {\doibase
  10.1038/s41567-020-0868-y} {\bibfield  {journal} {\bibinfo  {journal} {Nature
  Physics}\ }\textbf {\bibinfo {volume} {16}},\ \bibinfo {pages} {620--624}
  (\bibinfo {year} {2020})}\BibitemShut {NoStop}%
\bibitem [{\citenamefont {Henderson}\ \emph {et~al.}(2017)\citenamefont
  {Henderson} \emph {et~al.}}]{Henderson2018}%
  \BibitemOpen
  \bibfield  {author} {\bibinfo {author} {\bibfnamefont {J.}~\bibnamefont
  {Henderson}} \emph {et~al.},\ }\bibfield  {title} {\enquote {\bibinfo {title}
  {Testing microscopically derived descriptions of nuclear collectivity:
  Coulomb excitation of $^{22}$\text{Mg}},}\ }\href {\doibase
  10.1016/j.physletb.2018.05.064} {\bibfield  {journal} {\bibinfo  {journal}
  {Physics Letters B}\ }\textbf {\bibinfo {volume} {782}} (\bibinfo {year}
  {2017}),\ 10.1016/j.physletb.2018.05.064}\BibitemShut {NoStop}%
\bibitem [{\citenamefont {Henderson}\ \emph {et~al.}(2020)\citenamefont
  {Henderson} \emph {et~al.}}]{Henderson2020}%
  \BibitemOpen
  \bibfield  {author} {\bibinfo {author} {\bibfnamefont {J.}~\bibnamefont
  {Henderson}} \emph {et~al.},\ }\bibfield  {title} {\enquote {\bibinfo {title}
  {{Coulomb excitation of the $\left|T_z\right|=\frac{1}{2}$, $A=23$ mirror
  pair and systematics of ab-initio $E2$ strength}},}\ }\href@noop {} {\
  (\bibinfo {year} {2020})},\ \Eprint {http://arxiv.org/abs/2005.03796}
  {arXiv:2005.03796} \BibitemShut {NoStop}%
\bibitem [{\citenamefont {Minamisono}\ \emph {et~al.}(1981)\citenamefont
  {Minamisono}, \citenamefont {Nojiri},\ and\ \citenamefont
  {Ochi}}]{Minamisono1981}%
  \BibitemOpen
  \bibfield  {author} {\bibinfo {author} {\bibfnamefont {T.}~\bibnamefont
  {Minamisono}}, \bibinfo {author} {\bibfnamefont {Y.}~\bibnamefont {Nojiri}},
  \ and\ \bibinfo {author} {\bibfnamefont {S.}~\bibnamefont {Ochi}},\
  }\bibfield  {title} {\enquote {\bibinfo {title} {Measurement of the magnetic
  moment of the short-lived beta-emitter $^{28}$\text{Al} polarized by means of
  the overhauser effect in \text{Li} metal},}\ }\href {\doibase
  https://doi.org/10.1016/0370-2693(81)91075-3} {\bibfield  {journal} {\bibinfo
   {journal} {Physics Letters B}\ }\textbf {\bibinfo {volume} {106}},\ \bibinfo
  {pages} {38 -- 41} (\bibinfo {year} {1981})}\BibitemShut {NoStop}%
\bibitem [{\citenamefont {Morton}\ \emph {et~al.}(2002)\citenamefont {Morton}
  \emph {et~al.}}]{Morton2002}%
  \BibitemOpen
  \bibfield  {author} {\bibinfo {author} {\bibfnamefont {A.C}\ \bibnamefont
  {Morton}} \emph {et~al.},\ }\bibfield  {title} {\enquote {\bibinfo {title}
  {Beta decay studies of nuclei near $^{32}$\text{Mg}: Investigating the $\nu
  (f_{7/2})–(d_{3/2})$ inversion at the $\textit{N} = 20$ shell closure},}\
  }\href {\doibase http://dx.doi.org/10.1016/S0370-2693(02)02544-3} {\bibfield
  {journal} {\bibinfo  {journal} {Physics Letters B}\ }\textbf {\bibinfo
  {volume} {544}},\ \bibinfo {pages} {274 -- 279} (\bibinfo {year}
  {2002})}\BibitemShut {NoStop}%
\bibitem [{\citenamefont {Himpe}\ \emph {et~al.}(2006)\citenamefont {Himpe}
  \emph {et~al.}}]{Himpe2006}%
  \BibitemOpen
  \bibfield  {author} {\bibinfo {author} {\bibfnamefont {P.}~\bibnamefont
  {Himpe}} \emph {et~al.},\ }\bibfield  {title} {\enquote {\bibinfo {title} {g
  factors of $^{31,32,33}$\text{Al}: Indication for intruder configurations in
  the $^{33}$\text{Al} ground state},}\ }\href@noop {} {\bibfield  {journal}
  {\bibinfo  {journal} {Physics Letters B}\ }\textbf {\bibinfo {volume}
  {643}},\ \bibinfo {pages} {257 -- 262} (\bibinfo {year} {2006})}\BibitemShut
  {NoStop}%
\bibitem [{\citenamefont {Kameda}\ \emph {et~al.}(2007)\citenamefont {Kameda}
  \emph {et~al.}}]{Kameda2007}%
  \BibitemOpen
  \bibfield  {author} {\bibinfo {author} {\bibfnamefont {D.}~\bibnamefont
  {Kameda}} \emph {et~al.},\ }\bibfield  {title} {\enquote {\bibinfo {title}
  {Measurement of the electric quadrupole moment of $^{32}$\text{Al}},}\ }\href
  {\doibase http://dx.doi.org/10.1016/j.physletb.2007.01.063} {\bibfield
  {journal} {\bibinfo  {journal} {Physics Letters B}\ }\textbf {\bibinfo
  {volume} {647}},\ \bibinfo {pages} {93 -- 97} (\bibinfo {year}
  {2007})}\BibitemShut {NoStop}%
\bibitem [{\citenamefont {Kwiatkowski}\ \emph {et~al.}(2015)\citenamefont
  {Kwiatkowski} \emph {et~al.}}]{Kwiatkowski2015}%
  \BibitemOpen
  \bibfield  {author} {\bibinfo {author} {\bibfnamefont {A.~A.}\ \bibnamefont
  {Kwiatkowski}} \emph {et~al.},\ }\bibfield  {title} {\enquote {\bibinfo
  {title} {Observation of a crossover of ${S}_{2n}$ in the island of inversion
  from precision mass spectrometry},}\ }\href {\doibase
  10.1103/PhysRevC.92.061301} {\bibfield  {journal} {\bibinfo  {journal}
  {Physical Review C}\ }\textbf {\bibinfo {volume} {92}},\ \bibinfo {pages}
  {061301} (\bibinfo {year} {2015})}\BibitemShut {NoStop}%
\bibitem [{\citenamefont {Heylen}\ \emph {et~al.}(2016)\citenamefont {Heylen}
  \emph {et~al.}}]{Heylen2016}%
  \BibitemOpen
  \bibfield  {author} {\bibinfo {author} {\bibfnamefont {H.}~\bibnamefont
  {Heylen}} \emph {et~al.},\ }\bibfield  {title} {\enquote {\bibinfo {title}
  {High-precision quadrupole moment reveals significant intruder component in
  $_{13}^{33}\mathrm{Al}_{20}$ ground state},}\ }\href {\doibase
  10.1103/PhysRevC.94.034312} {\bibfield  {journal} {\bibinfo  {journal}
  {Physical Review C}\ }\textbf {\bibinfo {volume} {94}},\ \bibinfo {pages}
  {034312} (\bibinfo {year} {2016})}\BibitemShut {NoStop}%
\bibitem [{\citenamefont {Xu}\ \emph {et~al.}(2018)\citenamefont {Xu} \emph
  {et~al.}}]{Xu2018}%
  \BibitemOpen
  \bibfield  {author} {\bibinfo {author} {\bibfnamefont {Z.Y.}\ \bibnamefont
  {Xu}} \emph {et~al.},\ }\bibfield  {title} {\enquote {\bibinfo {title}
  {Nuclear moments of the low-lying isomeric $1^+$ state of ${}^{34}$\text{Al}:
  Investigation on the neutron 1p1h excitation across \textit{N} = 20 in the
  island of inversion},}\ }\href {\doibase
  https://doi.org/10.1016/j.physletb.2018.06.009} {\bibfield  {journal}
  {\bibinfo  {journal} {Physics Letters B}\ }\textbf {\bibinfo {volume}
  {782}},\ \bibinfo {pages} {619 -- 626} (\bibinfo {year} {2018})}\BibitemShut
  {NoStop}%
\bibitem [{\citenamefont {Neugart}\ \emph {et~al.}(2017)\citenamefont
  {Neugart}, \citenamefont {Billowes}, \citenamefont {Bissell}, \citenamefont
  {Blaum}, \citenamefont {Cheal}, \citenamefont {Flanagan}, \citenamefont
  {Neyens}, \citenamefont {N{\"o}rtersh{\"a}user},\ and\ \citenamefont
  {Yordanov}}]{Neugart2017}%
  \BibitemOpen
  \bibfield  {author} {\bibinfo {author} {\bibfnamefont {R}~\bibnamefont
  {Neugart}}, \bibinfo {author} {\bibfnamefont {J}~\bibnamefont {Billowes}},
  \bibinfo {author} {\bibfnamefont {M~L}\ \bibnamefont {Bissell}}, \bibinfo
  {author} {\bibfnamefont {K}~\bibnamefont {Blaum}}, \bibinfo {author}
  {\bibfnamefont {B}~\bibnamefont {Cheal}}, \bibinfo {author} {\bibfnamefont
  {K~T}\ \bibnamefont {Flanagan}}, \bibinfo {author} {\bibfnamefont
  {G}~\bibnamefont {Neyens}}, \bibinfo {author} {\bibfnamefont {W}~\bibnamefont
  {N{\"o}rtersh{\"a}user}}, \ and\ \bibinfo {author} {\bibfnamefont {D~T}\
  \bibnamefont {Yordanov}},\ }\bibfield  {title} {\enquote {\bibinfo {title}
  {Collinear laser spectroscopy at {ISOLDE}: new methods and highlights},}\
  }\href {\doibase 10.1088/1361-6471/aa6642} {\bibfield  {journal} {\bibinfo
  {journal} {Journal of Physics G: Nuclear and Particle Physics}\ }\textbf
  {\bibinfo {volume} {44}},\ \bibinfo {pages} {064002} (\bibinfo {year}
  {2017})}\BibitemShut {NoStop}%
\bibitem [{\citenamefont {Borge}\ and\ \citenamefont
  {Jonson}(2017)}]{Borge2017}%
  \BibitemOpen
  \bibfield  {author} {\bibinfo {author} {\bibfnamefont {M.J.G.}\ \bibnamefont
  {Borge}}\ and\ \bibinfo {author} {\bibfnamefont {B.J.}\ \bibnamefont
  {Jonson}},\ }\bibfield  {title} {\enquote {\bibinfo {title} {{ISOLDE past,
  present and future}},}\ }\href {\doibase 10.1088/1361-6471/aa5f03} {\bibfield
   {journal} {\bibinfo  {journal} {Journal of Physics. G: Nuclear and Particle
  Physics}\ }\textbf {\bibinfo {volume} {44}},\ \bibinfo {pages} {044011}
  (\bibinfo {year} {2017})},\ \bibinfo {note} {[Erratum: J.Phys.G 44, 079501
  (2017)]}\BibitemShut {NoStop}%
\bibitem [{\citenamefont {Kreim}\ \emph {et~al.}(2014)\citenamefont {Kreim}
  \emph {et~al.}}]{Kreim2014}%
  \BibitemOpen
  \bibfield  {author} {\bibinfo {author} {\bibfnamefont {K.}~\bibnamefont
  {Kreim}} \emph {et~al.},\ }\bibfield  {title} {\enquote {\bibinfo {title}
  {Nuclear charge radii of potassium isotopes beyond \textit{N} = 28},}\ }\href
  {\doibase http://dx.doi.org/10.1016/j.physletb.2014.02.012} {\bibfield
  {journal} {\bibinfo  {journal} {Physics Letters B}\ }\textbf {\bibinfo
  {volume} {731}},\ \bibinfo {pages} {97 -- 102} (\bibinfo {year}
  {2014})}\BibitemShut {NoStop}%
\bibitem [{\citenamefont {K\"oster}\ \emph {et~al.}(2003)\citenamefont
  {K\"oster} \emph {et~al.}}]{Koster2003}%
  \BibitemOpen
  \bibfield  {author} {\bibinfo {author} {\bibfnamefont {U.}~\bibnamefont
  {K\"oster}} \emph {et~al.} (\bibinfo {collaboration} {IS365, IS387, IS393 and
  ISOLDE Collaborations}),\ }\bibfield  {title} {\enquote {\bibinfo {title}
  {On-line yields obtained with the \text{ISOLDE RILIS}},}\ }\href {\doibase
  https://doi.org/10.1016/S0168-583X(02)01956-0} {\bibfield  {journal}
  {\bibinfo  {journal} {Nuclear Instruments and Methods in Physics Research
  Section B: Beam Interactions with Materials and Atoms}\ }\textbf {\bibinfo
  {volume} {204}},\ \bibinfo {pages} {347 -- 352} (\bibinfo {year} {2003})},\
  \bibinfo {note} {14th International Conference on Electromagnetic Isotope
  Separators and Techniques Related to their Applications}\BibitemShut
  {NoStop}%
\bibitem [{\citenamefont {Fr{\"a}nberg}\ \emph {et~al.}(2008)\citenamefont
  {Fr{\"a}nberg} \emph {et~al.}}]{Franberg2008}%
  \BibitemOpen
  \bibfield  {author} {\bibinfo {author} {\bibfnamefont {H.}~\bibnamefont
  {Fr{\"a}nberg}} \emph {et~al.},\ }\bibfield  {title} {\enquote {\bibinfo
  {title} {Off-line commissioning of the \text{ISOLDE} cooler},}\ }\href
  {\doibase https://doi.org/10.1016/j.nimb.2008.05.097} {\bibfield  {journal}
  {\bibinfo  {journal} {Nuclear Instruments and Methods in Physics Research
  Section B: Beam Interactions with Materials and Atoms}\ }\textbf {\bibinfo
  {volume} {266}},\ \bibinfo {pages} {4502 -- 4504} (\bibinfo {year} {2008})},\
  \bibinfo {note} {proceedings of the 15th International Conference on
  Electromagnetic Isotope Separators and Techniques Related to their
  Applications}\BibitemShut {NoStop}%
\bibitem [{\citenamefont {Gins}\ \emph {et~al.}(2018)\citenamefont {Gins},
  \citenamefont {de~Groote}, \citenamefont {Bissell}, \citenamefont {Buitrago},
  \citenamefont {Ferrer}, \citenamefont {Lynch}, \citenamefont {Neyens},\ and\
  \citenamefont {Sels}}]{Gins2018}%
  \BibitemOpen
  \bibfield  {author} {\bibinfo {author} {\bibfnamefont {W.}~\bibnamefont
  {Gins}}, \bibinfo {author} {\bibfnamefont {R.P.}\ \bibnamefont {de~Groote}},
  \bibinfo {author} {\bibfnamefont {M.L.}\ \bibnamefont {Bissell}}, \bibinfo
  {author} {\bibfnamefont {C.~Granados}\ \bibnamefont {Buitrago}}, \bibinfo
  {author} {\bibfnamefont {R.}~\bibnamefont {Ferrer}}, \bibinfo {author}
  {\bibfnamefont {K.M.}\ \bibnamefont {Lynch}}, \bibinfo {author}
  {\bibfnamefont {G.}~\bibnamefont {Neyens}}, \ and\ \bibinfo {author}
  {\bibfnamefont {S.}~\bibnamefont {Sels}},\ }\bibfield  {title} {\enquote
  {\bibinfo {title} {Analysis of counting data: Development of the satlas
  python package},}\ }\href {\doibase
  https://doi.org/10.1016/j.cpc.2017.09.012} {\bibfield  {journal} {\bibinfo
  {journal} {Computer Physics Communications}\ }\textbf {\bibinfo {volume}
  {222}},\ \bibinfo {pages} {286 -- 294} (\bibinfo {year} {2018})}\BibitemShut
  {NoStop}%
\bibitem [{\citenamefont {Bendali}\ \emph {et~al.}(1986)\citenamefont
  {Bendali}, \citenamefont {Duong}, \citenamefont {Juncar}, \citenamefont
  {Saint~Jalm},\ and\ \citenamefont {Vialle}}]{Bendali1986}%
  \BibitemOpen
  \bibfield  {author} {\bibinfo {author} {\bibfnamefont {N.}~\bibnamefont
  {Bendali}}, \bibinfo {author} {\bibfnamefont {H.~T.}\ \bibnamefont {Duong}},
  \bibinfo {author} {\bibfnamefont {P.}~\bibnamefont {Juncar}}, \bibinfo
  {author} {\bibfnamefont {J.~M.}\ \bibnamefont {Saint~Jalm}}, \ and\ \bibinfo
  {author} {\bibfnamefont {J.L.}\ \bibnamefont {Vialle}},\ }\bibfield  {title}
  {\enquote {\bibinfo {title} {Na$^{+}$-\text{Na} charge exchange processes
  studied by collinear laser spectroscopy},}\ }\href@noop {} {\bibfield
  {journal} {\bibinfo  {journal} {Journal of Physics B: Atomic, Molecular and
  Optical Physics}\ }\textbf {\bibinfo {volume} {19}},\ \bibinfo {pages} {233}
  (\bibinfo {year} {1986})}\BibitemShut {NoStop}%
\bibitem [{\citenamefont {Klose}\ \emph {et~al.}(2012)\citenamefont {Klose}
  \emph {et~al.}}]{Klose2012114}%
  \BibitemOpen
  \bibfield  {author} {\bibinfo {author} {\bibfnamefont {A.}~\bibnamefont
  {Klose}} \emph {et~al.},\ }\bibfield  {title} {\enquote {\bibinfo {title}
  {Tests of atomic charge-exchange cells for collinear laser spectroscopy},}\
  }\href {\doibase https://doi.org/10.1016/j.nima.2012.03.006} {\bibfield
  {journal} {\bibinfo  {journal} {Nuclear Instruments and Methods in Physics
  Research Section A: Accelerators, Spectrometers, Detectors and Associated
  Equipment}\ }\textbf {\bibinfo {volume} {678}},\ \bibinfo {pages} {114 --
  121} (\bibinfo {year} {2012})}\BibitemShut {NoStop}%
\bibitem [{\citenamefont {Lew}(1949)}]{Lew1949}%
  \BibitemOpen
  \bibfield  {author} {\bibinfo {author} {\bibfnamefont {H.}~\bibnamefont
  {Lew}},\ }\bibfield  {title} {\enquote {\bibinfo {title} {The hyperfine
  structure of the $^{2}\text{P}_{3/2}$ state of $^{27}$\text{Al}: The nuclear
  electric quadrupole moment},}\ }\href {\doibase 10.1103/PhysRev.76.1086}
  {\bibfield  {journal} {\bibinfo  {journal} {Physical Review}\ }\textbf
  {\bibinfo {volume} {76}},\ \bibinfo {pages} {1086--1092} (\bibinfo {year}
  {1949})}\BibitemShut {NoStop}%
\bibitem [{\citenamefont {Weber}\ \emph {et~al.}(1987)\citenamefont {Weber},
  \citenamefont {Lawrenz}, \citenamefont {Obrebski},\ and\ \citenamefont
  {Niemax}}]{Weber1987}%
  \BibitemOpen
  \bibfield  {author} {\bibinfo {author} {\bibfnamefont {K-H}\ \bibnamefont
  {Weber}}, \bibinfo {author} {\bibfnamefont {J}~\bibnamefont {Lawrenz}},
  \bibinfo {author} {\bibfnamefont {A}~\bibnamefont {Obrebski}}, \ and\
  \bibinfo {author} {\bibfnamefont {K}~\bibnamefont {Niemax}},\ }\bibfield
  {title} {\enquote {\bibinfo {title} {High-resolution laser spectroscopy of
  aluminium, gallium and thallium},}\ }\href {\doibase
  10.1088/0031-8949/35/3/014} {\bibfield  {journal} {\bibinfo  {journal}
  {Physica Scripta}\ }\textbf {\bibinfo {volume} {35}},\ \bibinfo {pages}
  {309--312} (\bibinfo {year} {1987})}\BibitemShut {NoStop}%
\bibitem [{\citenamefont {Levins}\ \emph {et~al.}(1997)\citenamefont {Levins},
  \citenamefont {Billowes}, \citenamefont {Campbell},\ and\ \citenamefont
  {Pearson}}]{Levins1997}%
  \BibitemOpen
  \bibfield  {author} {\bibinfo {author} {\bibfnamefont {J.M.G}\ \bibnamefont
  {Levins}}, \bibinfo {author} {\bibfnamefont {J.}~\bibnamefont {Billowes}},
  \bibinfo {author} {\bibfnamefont {P.}~\bibnamefont {Campbell}}, \ and\
  \bibinfo {author} {\bibfnamefont {M.R.}\ \bibnamefont {Pearson}},\ }\bibfield
   {title} {\enquote {\bibinfo {title} {The quadrupole moment of \text{Al}},}\
  }\href {\doibase 10.1088/0954-3899/23/9/015} {\bibfield  {journal} {\bibinfo
  {journal} {Journal of Physics G: Nuclear and Particle Physics}\ }\textbf
  {\bibinfo {volume} {23}},\ \bibinfo {pages} {1145--1149} (\bibinfo {year}
  {1997})}\BibitemShut {NoStop}%
\bibitem [{\citenamefont {Zhan-Kui}\ \emph {et~al.}(1982)\citenamefont
  {Zhan-Kui}, \citenamefont {Lundberg},\ and\ \citenamefont
  {Svanberg}}]{Jiang1982}%
  \BibitemOpen
  \bibfield  {author} {\bibinfo {author} {\bibfnamefont {Jiang}\ \bibnamefont
  {Zhan-Kui}}, \bibinfo {author} {\bibfnamefont {H.}~\bibnamefont {Lundberg}},
  \ and\ \bibinfo {author} {\bibfnamefont {S.}~\bibnamefont {Svanberg}},\
  }\bibfield  {title} {\enquote {\bibinfo {title} {Hyperfine structure of the
  4s $^2\text{S}_{1/2}$ state of $^{27}\text{Al}$},}\ }\href {\doibase
  https://doi.org/10.1016/0375-9601(82)90732-0} {\bibfield  {journal} {\bibinfo
   {journal} {Physics Letters A}\ }\textbf {\bibinfo {volume} {92}},\ \bibinfo
  {pages} {27 -- 28} (\bibinfo {year} {1982})}\BibitemShut {NoStop}%
\bibitem [{\citenamefont {Antusek}\ and\ \citenamefont
  {Holka}(2015)}]{Antusek2015}%
  \BibitemOpen
  \bibfield  {author} {\bibinfo {author} {\bibfnamefont {A.}~\bibnamefont
  {Antusek}}\ and\ \bibinfo {author} {\bibfnamefont {F.}~\bibnamefont
  {Holka}},\ }\bibfield  {title} {\enquote {\bibinfo {title} {Absolute
  shielding scales for \text{Al, Ga, and In} and revised nuclear magnetic
  dipole moments of $^{27}$\text{Al}, $^{69}$\text{Ga}, $^{71}$\text{Ga},
  $^{113}$\text{In}, and $^{115}$\text{In} nuclei},}\ }\href {\doibase
  10.1063/1.4928592} {\bibfield  {journal} {\bibinfo  {journal} {The Journal of
  Chemical Physics}\ }\textbf {\bibinfo {volume} {143}},\ \bibinfo {pages}
  {074301} (\bibinfo {year} {2015})}\BibitemShut {NoStop}%
\bibitem [{\citenamefont {Kell{\"o}}\ \emph {et~al.}(1999)\citenamefont
  {Kell{\"o}}, \citenamefont {Sadlej}, \citenamefont {Pyykk{\"o}},
  \citenamefont {Sundholm},\ and\ \citenamefont {Tokman}}]{Kello1999}%
  \BibitemOpen
  \bibfield  {author} {\bibinfo {author} {\bibfnamefont {V.}~\bibnamefont
  {Kell{\"o}}}, \bibinfo {author} {\bibfnamefont {A.~J.}\ \bibnamefont
  {Sadlej}}, \bibinfo {author} {\bibfnamefont {P.}~\bibnamefont {Pyykk{\"o}}},
  \bibinfo {author} {\bibfnamefont {D.}~\bibnamefont {Sundholm}}, \ and\
  \bibinfo {author} {\bibfnamefont {M.}~\bibnamefont {Tokman}},\ }\bibfield
  {title} {\enquote {\bibinfo {title} {Electric quadrupole moment of the
  $^{27}$\text{Al} nucleus: Converging results from the \text{AlF} and
  \text{AlCl} molecules and the \text{Al} atom},}\ }\href {\doibase
  http://dx.doi.org/10.1016/S0009-2614(99)00340-1} {\bibfield  {journal}
  {\bibinfo  {journal} {Chemical Physics Letters}\ }\textbf {\bibinfo {volume}
  {304}},\ \bibinfo {pages} {414 -- 422} (\bibinfo {year} {1999})}\BibitemShut
  {NoStop}%
\bibitem [{\citenamefont {Stone}(2019)}]{Stone2019}%
  \BibitemOpen
  \bibfield  {author} {\bibinfo {author} {\bibfnamefont {N.J.}\ \bibnamefont
  {Stone}},\ }\href@noop {} {\emph {\bibinfo {title} {Table of recommended
  nuclear magnetic dipole moments}}},\ \bibinfo {type} {Tech. Rep.}\ (\bibinfo
  {institution} {INDC(NDS)-0794},\ \bibinfo {year} {2019})\BibitemShut
  {NoStop}%
\bibitem [{\citenamefont {De~Rydt}\ \emph {et~al.}(2013)\citenamefont
  {De~Rydt}, \citenamefont {Depuydt},\ and\ \citenamefont
  {Neyens}}]{Derydt2013}%
  \BibitemOpen
  \bibfield  {author} {\bibinfo {author} {\bibfnamefont {M.}~\bibnamefont
  {De~Rydt}}, \bibinfo {author} {\bibfnamefont {M.}~\bibnamefont {Depuydt}}, \
  and\ \bibinfo {author} {\bibfnamefont {G.}~\bibnamefont {Neyens}},\
  }\bibfield  {title} {\enquote {\bibinfo {title} {Evaluation of the
  ground-state quadrupole moments of the (sd) nuclei},}\ }\href {\doibase
  http://dx.doi.org/10.1016/j.adt.2011.12.005} {\bibfield  {journal} {\bibinfo
  {journal} {Atomic Data and Nuclear Data Tables}\ }\textbf {\bibinfo {volume}
  {99}},\ \bibinfo {pages} {391 -- 415} (\bibinfo {year} {2013})}\BibitemShut
  {NoStop}%
\bibitem [{\citenamefont {Martensson-Pendrill}\ \emph
  {et~al.}(1990)\citenamefont {Martensson-Pendrill}, \citenamefont {Pendrill},
  \citenamefont {Salomonson}, \citenamefont {Ynnerman},\ and\ \citenamefont
  {Warston}}]{MartenssonPendrill1990}%
  \BibitemOpen
  \bibfield  {author} {\bibinfo {author} {\bibfnamefont {A~M}\ \bibnamefont
  {Martensson-Pendrill}}, \bibinfo {author} {\bibfnamefont {L}~\bibnamefont
  {Pendrill}}, \bibinfo {author} {\bibfnamefont {A}~\bibnamefont {Salomonson}},
  \bibinfo {author} {\bibfnamefont {A}~\bibnamefont {Ynnerman}}, \ and\
  \bibinfo {author} {\bibfnamefont {H}~\bibnamefont {Warston}},\ }\bibfield
  {title} {\enquote {\bibinfo {title} {Reanalysis of the isotope shift and
  nuclear charge radii in radioactive potassium isotopes},}\ }\href {\doibase
  10.1088/0953-4075/23/11/012} {\bibfield  {journal} {\bibinfo  {journal}
  {Journal of Physics B: Atomic, Molecular and Optical Physics}\ }\textbf
  {\bibinfo {volume} {23}},\ \bibinfo {pages} {1749--1761} (\bibinfo {year}
  {1990})}\BibitemShut {NoStop}%
\bibitem [{\citenamefont {Cheal}\ \emph {et~al.}(2012)\citenamefont {Cheal},
  \citenamefont {Cocolios},\ and\ \citenamefont {Fritzsche}}]{Cheal2012}%
  \BibitemOpen
  \bibfield  {author} {\bibinfo {author} {\bibfnamefont {B.}~\bibnamefont
  {Cheal}}, \bibinfo {author} {\bibfnamefont {T.~E.}\ \bibnamefont {Cocolios}},
  \ and\ \bibinfo {author} {\bibfnamefont {S.}~\bibnamefont {Fritzsche}},\
  }\bibfield  {title} {\enquote {\bibinfo {title} {Laser spectroscopy of
  radioactive isotopes: Role and limitations of accurate isotope-shift
  calculations},}\ }\href {\doibase 10.1103/PhysRevA.86.042501} {\bibfield
  {journal} {\bibinfo  {journal} {Physical Review A}\ }\textbf {\bibinfo
  {volume} {86}},\ \bibinfo {pages} {042501} (\bibinfo {year}
  {2012})}\BibitemShut {NoStop}%
\bibitem [{\citenamefont {N{\"o}rtersh{\"a}user}\ and\ \citenamefont
  {Geppert}(2014)}]{Noertershaeuser2014}%
  \BibitemOpen
  \bibfield  {author} {\bibinfo {author} {\bibfnamefont {W.}~\bibnamefont
  {N{\"o}rtersh{\"a}user}}\ and\ \bibinfo {author} {\bibfnamefont
  {C.}~\bibnamefont {Geppert}},\ }\enquote {\bibinfo {title} {Nuclear charge
  radii of light elements and recent developments in collinear laser
  spectroscopy},}\ in\ \href {\doibase 10.1007/978-3-642-45141-6_6} {\emph
  {\bibinfo {booktitle} {The Euroschool on Exotic Beams, Vol. IV}}},\ \bibinfo
  {editor} {edited by\ \bibinfo {editor} {\bibfnamefont {C.}~\bibnamefont
  {Scheidenberger}}\ and\ \bibinfo {editor} {\bibfnamefont {M.}~\bibnamefont
  {Pf{\"u}tzner}}}\ (\bibinfo  {publisher} {Springer Berlin Heidelberg},\
  \bibinfo {address} {Berlin, Heidelberg},\ \bibinfo {year} {2014})\ pp.\
  \bibinfo {pages} {233--292}\BibitemShut {NoStop}%
\bibitem [{\citenamefont {Filippin}\ \emph {et~al.}(2016)\citenamefont
  {Filippin}, \citenamefont {Beerwerth}, \citenamefont {Ekman}, \citenamefont
  {Fritzsche}, \citenamefont {Godefroid},\ and\ \citenamefont
  {J\"onsson}}]{Filippin2016}%
  \BibitemOpen
  \bibfield  {author} {\bibinfo {author} {\bibfnamefont {L.}~\bibnamefont
  {Filippin}}, \bibinfo {author} {\bibfnamefont {R.}~\bibnamefont {Beerwerth}},
  \bibinfo {author} {\bibfnamefont {J.}~\bibnamefont {Ekman}}, \bibinfo
  {author} {\bibfnamefont {S.}~\bibnamefont {Fritzsche}}, \bibinfo {author}
  {\bibfnamefont {M.}~\bibnamefont {Godefroid}}, \ and\ \bibinfo {author}
  {\bibfnamefont {P.}~\bibnamefont {J\"onsson}},\ }\bibfield  {title} {\enquote
  {\bibinfo {title} {Multiconfiguration calculations of electronic isotope
  shift factors in al i},}\ }\href {\doibase 10.1103/PhysRevA.94.062508}
  {\bibfield  {journal} {\bibinfo  {journal} {Physical Review A}\ }\textbf
  {\bibinfo {volume} {94}},\ \bibinfo {pages} {062508} (\bibinfo {year}
  {2016})}\BibitemShut {NoStop}%
\bibitem [{\citenamefont {Yordanov}\ \emph {et~al.}(2012)\citenamefont
  {Yordanov} \emph {et~al.}}]{Yordanov2012}%
  \BibitemOpen
  \bibfield  {author} {\bibinfo {author} {\bibfnamefont {D.~T.}\ \bibnamefont
  {Yordanov}} \emph {et~al.},\ }\bibfield  {title} {\enquote {\bibinfo {title}
  {Nuclear charge radii of
  $^{21\mathrm{\text{\ensuremath{-}}}32}\mathrm{Mg}$},}\ }\href {\doibase
  10.1103/PhysRevLett.108.042504} {\bibfield  {journal} {\bibinfo  {journal}
  {Physical Review Letters}\ }\textbf {\bibinfo {volume} {108}},\ \bibinfo
  {pages} {042504} (\bibinfo {year} {2012})}\BibitemShut {NoStop}%
\bibitem [{\citenamefont {Angeli}\ and\ \citenamefont
  {Marinova}(2013)}]{Angeli2013}%
  \BibitemOpen
  \bibfield  {author} {\bibinfo {author} {\bibfnamefont {I.}~\bibnamefont
  {Angeli}}\ and\ \bibinfo {author} {\bibfnamefont {K.P.}\ \bibnamefont
  {Marinova}},\ }\bibfield  {title} {\enquote {\bibinfo {title} {Table of
  experimental nuclear ground state charge radii: An update},}\ }\href
  {\doibase https://doi.org/10.1016/j.adt.2011.12.006} {\bibfield  {journal}
  {\bibinfo  {journal} {Atomic Data and Nuclear Data Tables}\ }\textbf
  {\bibinfo {volume} {99}},\ \bibinfo {pages} {69 -- 95} (\bibinfo {year}
  {2013})}\BibitemShut {NoStop}%
\bibitem [{\citenamefont {Blaum}\ \emph {et~al.}(2008)\citenamefont {Blaum},
  \citenamefont {Geithner}, \citenamefont {Lassen}, \citenamefont {Lievens},
  \citenamefont {Marinova},\ and\ \citenamefont {Neugart}}]{Blaum2008}%
  \BibitemOpen
  \bibfield  {author} {\bibinfo {author} {\bibfnamefont {K.}~\bibnamefont
  {Blaum}}, \bibinfo {author} {\bibfnamefont {W.}~\bibnamefont {Geithner}},
  \bibinfo {author} {\bibfnamefont {J.}~\bibnamefont {Lassen}}, \bibinfo
  {author} {\bibfnamefont {P.}~\bibnamefont {Lievens}}, \bibinfo {author}
  {\bibfnamefont {K.}~\bibnamefont {Marinova}}, \ and\ \bibinfo {author}
  {\bibfnamefont {R.}~\bibnamefont {Neugart}},\ }\bibfield  {title} {\enquote
  {\bibinfo {title} {{Nuclear moments and charge radii of argon isotopes
  between the neutron-shell closures N=20 and N=28}},}\ }\href {\doibase
  10.1016/j.nuclphysa.2007.11.004} {\bibfield  {journal} {\bibinfo  {journal}
  {Nuclear Physics A}\ }\textbf {\bibinfo {volume} {799}},\ \bibinfo {pages}
  {30--45} (\bibinfo {year} {2008})}\BibitemShut {NoStop}%
\bibitem [{\citenamefont {Fricke}\ and\ \citenamefont
  {Heilig}(2004)}]{Fricke2004}%
  \BibitemOpen
  \bibfield  {author} {\bibinfo {author} {\bibfnamefont {G.}~\bibnamefont
  {Fricke}}\ and\ \bibinfo {author} {\bibfnamefont {K.}~\bibnamefont
  {Heilig}},\ }\href {\doibase 10.1007/10856314_42} {\emph {\bibinfo {title}
  {Nuclear Charge Radii}}}\ (\bibinfo  {publisher} {Springer-Verlag Berlin
  Heidelberg},\ \bibinfo {year} {2004})\BibitemShut {NoStop}%
\bibitem [{\citenamefont {{De Vries}}\ \emph {et~al.}(1987)\citenamefont {{De
  Vries}}, \citenamefont {{De Jager}},\ and\ \citenamefont {{De
  Vries}}}]{Devries1987}%
  \BibitemOpen
  \bibfield  {author} {\bibinfo {author} {\bibfnamefont {H.}~\bibnamefont {{De
  Vries}}}, \bibinfo {author} {\bibfnamefont {C.W.}\ \bibnamefont {{De
  Jager}}}, \ and\ \bibinfo {author} {\bibfnamefont {C.}~\bibnamefont {{De
  Vries}}},\ }\bibfield  {title} {\enquote {\bibinfo {title} {Nuclear
  charge-density-distribution parameters from elastic electron scattering},}\
  }\href {\doibase https://doi.org/10.1016/0092-640X(87)90013-1} {\bibfield
  {journal} {\bibinfo  {journal} {Atomic Data and Nuclear Data Tables}\
  }\textbf {\bibinfo {volume} {36}},\ \bibinfo {pages} {495 -- 536} (\bibinfo
  {year} {1987})}\BibitemShut {NoStop}%
\bibitem [{\citenamefont {Lombard}\ and\ \citenamefont
  {Bishop}(1967)}]{Lombard1967}%
  \BibitemOpen
  \bibfield  {author} {\bibinfo {author} {\bibfnamefont {R.M.}\ \bibnamefont
  {Lombard}}\ and\ \bibinfo {author} {\bibfnamefont {G.R.}\ \bibnamefont
  {Bishop}},\ }\bibfield  {title} {\enquote {\bibinfo {title} {The scattering
  of high-energy electrons by $^{27}$\text{Al}},}\ }\href {\doibase
  https://doi.org/10.1016/0375-9474(67)90655-0} {\bibfield  {journal} {\bibinfo
   {journal} {Nuclear Physics A}\ }\textbf {\bibinfo {volume} {101}},\ \bibinfo
  {pages} {601 -- 624} (\bibinfo {year} {1967})}\BibitemShut {NoStop}%
\bibitem [{\citenamefont {{Fey}}\ \emph {et~al.}(1973)\citenamefont {{Fey}},
  \citenamefont {{Frank}}, \citenamefont {{Sch{\"u}tz}},\ and\ \citenamefont
  {{Theissen}}}]{Fey1973}%
  \BibitemOpen
  \bibfield  {author} {\bibinfo {author} {\bibfnamefont {G.}~\bibnamefont
  {{Fey}}}, \bibinfo {author} {\bibfnamefont {H.}~\bibnamefont {{Frank}}},
  \bibinfo {author} {\bibfnamefont {W.}~\bibnamefont {{Sch{\"u}tz}}}, \ and\
  \bibinfo {author} {\bibfnamefont {H.}~\bibnamefont {{Theissen}}},\ }\bibfield
   {title} {\enquote {\bibinfo {title} {{Nuclear Rms charge radii from relative
  electron scattering measurements at low energies}},}\ }\href {\doibase
  10.1007/BF01391614} {\bibfield  {journal} {\bibinfo  {journal} {Zeitschrift
  fur Physik}\ }\textbf {\bibinfo {volume} {265}},\ \bibinfo {pages} {401--403}
  (\bibinfo {year} {1973})}\BibitemShut {NoStop}%
\bibitem [{\citenamefont {Rossi}\ \emph {et~al.}(2015)\citenamefont {Rossi}
  \emph {et~al.}}]{Rossi2015}%
  \BibitemOpen
  \bibfield  {author} {\bibinfo {author} {\bibfnamefont {D.~M.}\ \bibnamefont
  {Rossi}} \emph {et~al.},\ }\bibfield  {title} {\enquote {\bibinfo {title}
  {Charge radii of neutron-deficient $^{36}\mathrm{K}$ and
  $^{37}\mathrm{K}$},}\ }\href {\doibase 10.1103/PhysRevC.92.014305} {\bibfield
   {journal} {\bibinfo  {journal} {Physical Review C}\ }\textbf {\bibinfo
  {volume} {92}},\ \bibinfo {pages} {014305} (\bibinfo {year}
  {2015})}\BibitemShut {NoStop}%
\bibitem [{\citenamefont {Huber}\ \emph {et~al.}(1978)\citenamefont {Huber}
  \emph {et~al.}}]{Huber1978}%
  \BibitemOpen
  \bibfield  {author} {\bibinfo {author} {\bibfnamefont {G.}~\bibnamefont
  {Huber}} \emph {et~al.},\ }\bibfield  {title} {\enquote {\bibinfo {title}
  {Spins, magnetic moments, and isotope shifts of $^{21-31}\mathrm{Na}$ by high
  resolution laser spectroscopy of the atomic ${D}_{1}$ line},}\ }\href
  {\doibase 10.1103/PhysRevC.18.2342} {\bibfield  {journal} {\bibinfo
  {journal} {Physical Review C}\ }\textbf {\bibinfo {volume} {18}},\ \bibinfo
  {pages} {2342--2354} (\bibinfo {year} {1978})}\BibitemShut {NoStop}%
\bibitem [{\citenamefont {Touchard}\ \emph {et~al.}(1982)\citenamefont
  {Touchard} \emph {et~al.}}]{Touchard1982}%
  \BibitemOpen
  \bibfield  {author} {\bibinfo {author} {\bibfnamefont {F.}~\bibnamefont
  {Touchard}} \emph {et~al.},\ }\bibfield  {title} {\enquote {\bibinfo {title}
  {Electric quadrupole moments and isotope shifts of radioactive sodium
  isotopes},}\ }\href {\doibase 10.1103/PhysRevC.25.2756} {\bibfield  {journal}
  {\bibinfo  {journal} {Physical Review C}\ }\textbf {\bibinfo {volume} {25}},\
  \bibinfo {pages} {2756--2770} (\bibinfo {year} {1982})}\BibitemShut {NoStop}%
\bibitem [{\citenamefont {Morris}\ \emph {et~al.}(2015)\citenamefont {Morris},
  \citenamefont {Parzuchowski},\ and\ \citenamefont {Bogner}}]{Morris2015}%
  \BibitemOpen
  \bibfield  {author} {\bibinfo {author} {\bibfnamefont {T.~D.}\ \bibnamefont
  {Morris}}, \bibinfo {author} {\bibfnamefont {N.~M.}\ \bibnamefont
  {Parzuchowski}}, \ and\ \bibinfo {author} {\bibfnamefont {S.~K.}\
  \bibnamefont {Bogner}},\ }\bibfield  {title} {\enquote {\bibinfo {title}
  {Magnus expansion and in-medium similarity renormalization group},}\ }\href
  {\doibase 10.1103/PhysRevC.92.034331} {\bibfield  {journal} {\bibinfo
  {journal} {Physical Review C}\ }\textbf {\bibinfo {volume} {92}},\ \bibinfo
  {pages} {034331} (\bibinfo {year} {2015})}\BibitemShut {NoStop}%
\bibitem [{\citenamefont {Simonis}\ \emph {et~al.}(2017)\citenamefont
  {Simonis}, \citenamefont {Stroberg}, \citenamefont {Hebeler}, \citenamefont
  {Holt},\ and\ \citenamefont {Schwenk}}]{Simonis2017}%
  \BibitemOpen
  \bibfield  {author} {\bibinfo {author} {\bibfnamefont {J.}~\bibnamefont
  {Simonis}}, \bibinfo {author} {\bibfnamefont {S.~R.}\ \bibnamefont
  {Stroberg}}, \bibinfo {author} {\bibfnamefont {K.}~\bibnamefont {Hebeler}},
  \bibinfo {author} {\bibfnamefont {J.~D.}\ \bibnamefont {Holt}}, \ and\
  \bibinfo {author} {\bibfnamefont {A.}~\bibnamefont {Schwenk}},\ }\bibfield
  {title} {\enquote {\bibinfo {title} {Saturation with chiral interactions and
  consequences for finite nuclei},}\ }\href {\doibase
  10.1103/PhysRevC.96.014303} {\bibfield  {journal} {\bibinfo  {journal}
  {Physical Review C}\ }\textbf {\bibinfo {volume} {96}},\ \bibinfo {pages}
  {014303} (\bibinfo {year} {2017})}\BibitemShut {NoStop}%
\bibitem [{\citenamefont {Parzuchowski}\ \emph {et~al.}(2017)\citenamefont
  {Parzuchowski}, \citenamefont {Stroberg}, \citenamefont {Navr\'atil},
  \citenamefont {Hergert},\ and\ \citenamefont {Bogner}}]{Parzuchowski2017}%
  \BibitemOpen
  \bibfield  {author} {\bibinfo {author} {\bibfnamefont {N.~M.}\ \bibnamefont
  {Parzuchowski}}, \bibinfo {author} {\bibfnamefont {S.~R.}\ \bibnamefont
  {Stroberg}}, \bibinfo {author} {\bibfnamefont {P.}~\bibnamefont
  {Navr\'atil}}, \bibinfo {author} {\bibfnamefont {H.}~\bibnamefont {Hergert}},
  \ and\ \bibinfo {author} {\bibfnamefont {S.~K.}\ \bibnamefont {Bogner}},\
  }\bibfield  {title} {\enquote {\bibinfo {title} {Ab initio electromagnetic
  observables with the in-medium similarity renormalization group},}\ }\href
  {\doibase 10.1103/PhysRevC.96.034324} {\bibfield  {journal} {\bibinfo
  {journal} {Physical Review C}\ }\textbf {\bibinfo {volume} {96}},\ \bibinfo
  {pages} {034324} (\bibinfo {year} {2017})}\BibitemShut {NoStop}%
\bibitem [{\citenamefont {Shimizu}\ \emph {et~al.}(2019)\citenamefont
  {Shimizu}, \citenamefont {Mizusaki}, \citenamefont {Utsuno},\ and\
  \citenamefont {Tsunoda}}]{Shimizu2019}%
  \BibitemOpen
  \bibfield  {author} {\bibinfo {author} {\bibfnamefont {N.}~\bibnamefont
  {Shimizu}}, \bibinfo {author} {\bibfnamefont {T.}~\bibnamefont {Mizusaki}},
  \bibinfo {author} {\bibfnamefont {Y.}~\bibnamefont {Utsuno}}, \ and\ \bibinfo
  {author} {\bibfnamefont {Y.}~\bibnamefont {Tsunoda}},\ }\bibfield  {title}
  {\enquote {\bibinfo {title} {Thick-restart block lanczos method for
  large-scale shell-model calculations},}\ }\href {\doibase
  https://doi.org/10.1016/j.cpc.2019.06.011} {\bibfield  {journal} {\bibinfo
  {journal} {Computer Physics Communications}\ }\textbf {\bibinfo {volume}
  {244}},\ \bibinfo {pages} {372 -- 384} (\bibinfo {year} {2019})}\BibitemShut
  {NoStop}%
\bibitem [{\citenamefont {Stroberg}()}]{imsrgplusplus}%
  \BibitemOpen
  \bibfield  {author} {\bibinfo {author} {\bibfnamefont {S.~R.}\ \bibnamefont
  {Stroberg}},\ }\href {https://github.com/ragnarstroberg/imsrg.} {\enquote
  {\bibinfo {title} {\url{https://github.com/ragnarstroberg/imsrg}},}\
  }\BibitemShut {NoStop}%
\bibitem [{\citenamefont {Hebeler}\ \emph {et~al.}(2011)\citenamefont
  {Hebeler}, \citenamefont {Bogner}, \citenamefont {Furnstahl}, \citenamefont
  {Nogga},\ and\ \citenamefont {Schwenk}}]{Hebeler2011}%
  \BibitemOpen
  \bibfield  {author} {\bibinfo {author} {\bibfnamefont {K.}~\bibnamefont
  {Hebeler}}, \bibinfo {author} {\bibfnamefont {S.~K.}\ \bibnamefont {Bogner}},
  \bibinfo {author} {\bibfnamefont {R.~J.}\ \bibnamefont {Furnstahl}}, \bibinfo
  {author} {\bibfnamefont {A.}~\bibnamefont {Nogga}}, \ and\ \bibinfo {author}
  {\bibfnamefont {A.}~\bibnamefont {Schwenk}},\ }\bibfield  {title} {\enquote
  {\bibinfo {title} {Improved nuclear matter calculations from chiral
  low-momentum interactions},}\ }\href {\doibase 10.1103/PhysRevC.83.031301}
  {\bibfield  {journal} {\bibinfo  {journal} {Physical Review C}\ }\textbf
  {\bibinfo {volume} {83}},\ \bibinfo {pages} {031301} (\bibinfo {year}
  {2011})}\BibitemShut {NoStop}%
\bibitem [{\citenamefont {Simonis}\ \emph {et~al.}(2016)\citenamefont
  {Simonis}, \citenamefont {Hebeler}, \citenamefont {Holt}, \citenamefont
  {Menendez},\ and\ \citenamefont {Schwenk}}]{Simonis2016}%
  \BibitemOpen
  \bibfield  {author} {\bibinfo {author} {\bibfnamefont {J.}~\bibnamefont
  {Simonis}}, \bibinfo {author} {\bibfnamefont {K.}~\bibnamefont {Hebeler}},
  \bibinfo {author} {\bibfnamefont {J.~D.}\ \bibnamefont {Holt}}, \bibinfo
  {author} {\bibfnamefont {J.}~\bibnamefont {Menendez}}, \ and\ \bibinfo
  {author} {\bibfnamefont {A.}~\bibnamefont {Schwenk}},\ }\bibfield  {title}
  {\enquote {\bibinfo {title} {{Exploring sd-shell nuclei from two- and
  three-nucleon interactions with realistic saturation properties}},}\ }\href
  {\doibase 10.1103/PhysRevC.93.011302} {\bibfield  {journal} {\bibinfo
  {journal} {Physical Review C}\ }\textbf {\bibinfo {volume} {93}},\ \bibinfo
  {pages} {011302} (\bibinfo {year} {2016})}\BibitemShut {NoStop}%
\bibitem [{\citenamefont {Ekstr\"om}\ \emph {et~al.}(2015)\citenamefont
  {Ekstr\"om}, \citenamefont {Jansen}, \citenamefont {Wendt}, \citenamefont
  {Hagen}, \citenamefont {Papenbrock}, \citenamefont {Carlsson}, \citenamefont
  {Forss\'en}, \citenamefont {Hjorth-Jensen}, \citenamefont {Navr\'atil},\ and\
  \citenamefont {Nazarewicz}}]{Eckstrom2015}%
  \BibitemOpen
  \bibfield  {author} {\bibinfo {author} {\bibfnamefont {A.}~\bibnamefont
  {Ekstr\"om}}, \bibinfo {author} {\bibfnamefont {G.~R.}\ \bibnamefont
  {Jansen}}, \bibinfo {author} {\bibfnamefont {K.~A.}\ \bibnamefont {Wendt}},
  \bibinfo {author} {\bibfnamefont {G.}~\bibnamefont {Hagen}}, \bibinfo
  {author} {\bibfnamefont {T.}~\bibnamefont {Papenbrock}}, \bibinfo {author}
  {\bibfnamefont {B.~D.}\ \bibnamefont {Carlsson}}, \bibinfo {author}
  {\bibfnamefont {C.}~\bibnamefont {Forss\'en}}, \bibinfo {author}
  {\bibfnamefont {M.}~\bibnamefont {Hjorth-Jensen}}, \bibinfo {author}
  {\bibfnamefont {P.}~\bibnamefont {Navr\'atil}}, \ and\ \bibinfo {author}
  {\bibfnamefont {W.}~\bibnamefont {Nazarewicz}},\ }\bibfield  {title}
  {\enquote {\bibinfo {title} {Accurate nuclear radii and binding energies from
  a chiral interaction},}\ }\href {\doibase 10.1103/PhysRevC.91.051301}
  {\bibfield  {journal} {\bibinfo  {journal} {Physical Review C}\ }\textbf
  {\bibinfo {volume} {91}},\ \bibinfo {pages} {051301} (\bibinfo {year}
  {2015})}\BibitemShut {NoStop}%
\bibitem [{\citenamefont {Garcia~Ruiz}\ \emph {et~al.}(2016)\citenamefont
  {Garcia~Ruiz} \emph {et~al.}}]{GarciaRuiz2016}%
  \BibitemOpen
  \bibfield  {author} {\bibinfo {author} {\bibfnamefont {R.~F.}\ \bibnamefont
  {Garcia~Ruiz}} \emph {et~al.},\ }\bibfield  {title} {\enquote {\bibinfo
  {title} {Unexpectedly large charge radii of neutron-rich calcium isotopes},}\
  }\href {\doibase http://dx.doi.org/10.1038/nphys3645} {\bibfield  {journal}
  {\bibinfo  {journal} {Nature Physics}\ }\textbf {\bibinfo {volume} {12}},\
  \bibinfo {pages} {594--598} (\bibinfo {year} {2016})}\BibitemShut {NoStop}%
\bibitem [{\citenamefont {Lapoux}\ \emph {et~al.}(2016)\citenamefont {Lapoux},
  \citenamefont {Som\`a}, \citenamefont {Barbieri}, \citenamefont {Hergert},
  \citenamefont {Holt},\ and\ \citenamefont {Stroberg}}]{Lapoux2016}%
  \BibitemOpen
  \bibfield  {author} {\bibinfo {author} {\bibfnamefont {V.}~\bibnamefont
  {Lapoux}}, \bibinfo {author} {\bibfnamefont {V.}~\bibnamefont {Som\`a}},
  \bibinfo {author} {\bibfnamefont {C.}~\bibnamefont {Barbieri}}, \bibinfo
  {author} {\bibfnamefont {H.}~\bibnamefont {Hergert}}, \bibinfo {author}
  {\bibfnamefont {J.~D.}\ \bibnamefont {Holt}}, \ and\ \bibinfo {author}
  {\bibfnamefont {S.~R.}\ \bibnamefont {Stroberg}},\ }\bibfield  {title}
  {\enquote {\bibinfo {title} {Radii and binding energies in oxygen isotopes: A
  challenge for nuclear forces},}\ }\href {\doibase
  10.1103/PhysRevLett.117.052501} {\bibfield  {journal} {\bibinfo  {journal}
  {Physical Review Letters}\ }\textbf {\bibinfo {volume} {117}},\ \bibinfo
  {pages} {052501} (\bibinfo {year} {2016})}\BibitemShut {NoStop}%
\bibitem [{\citenamefont {Hagen}\ \emph {et~al.}(2015)\citenamefont {Hagen}
  \emph {et~al.}}]{Hagen2015}%
  \BibitemOpen
  \bibfield  {author} {\bibinfo {author} {\bibfnamefont {G.}~\bibnamefont
  {Hagen}} \emph {et~al.},\ }\bibfield  {title} {\enquote {\bibinfo {title}
  {{Neutron and weak-charge distributions of the $^{48}$Ca nucleus}},}\
  }\href@noop {} {\bibfield  {journal} {\bibinfo  {journal} {Nature Phys.}\
  }\textbf {\bibinfo {volume} {12}},\ \bibinfo {pages} {186--190} (\bibinfo
  {year} {2015})}\BibitemShut {NoStop}%
\bibitem [{\citenamefont {Kaufmann}\ \emph {et~al.}(2020)\citenamefont
  {Kaufmann} \emph {et~al.}}]{Kaufmann2020}%
  \BibitemOpen
  \bibfield  {author} {\bibinfo {author} {\bibfnamefont {S.}~\bibnamefont
  {Kaufmann}} \emph {et~al.},\ }\bibfield  {title} {\enquote {\bibinfo {title}
  {Charge radius of the short-lived $^{68}\mathrm{Ni}$ and correlation with the
  dipole polarizability},}\ }\href {\doibase 10.1103/PhysRevLett.124.132502}
  {\bibfield  {journal} {\bibinfo  {journal} {Physical Review Letters}\
  }\textbf {\bibinfo {volume} {124}},\ \bibinfo {pages} {132502} (\bibinfo
  {year} {2020})}\BibitemShut {NoStop}%
\bibitem [{\citenamefont {Matsuta}\ \emph {et~al.}(2008)\citenamefont {Matsuta}
  \emph {et~al.}}]{Matsuta2007}%
  \BibitemOpen
  \bibfield  {author} {\bibinfo {author} {\bibfnamefont {K.}~\bibnamefont
  {Matsuta}} \emph {et~al.},\ }\bibfield  {title} {\enquote {\bibinfo {title}
  {Hyperfine interaction of 25\text{Al} in $\alpha$-\text{Al}$_2$\text{O}$_3$
  and its quadrupole moment},}\ }in\ \href@noop {} {\emph {\bibinfo {booktitle}
  {HFI/NQI 2007}}},\ \bibinfo {editor} {edited by\ \bibinfo {editor}
  {\bibfnamefont {A.}~\bibnamefont {Pasquevich}}, \bibinfo {editor}
  {\bibfnamefont {M.}~\bibnamefont {Renter{\'i}a}}, \bibinfo {editor}
  {\bibfnamefont {E.~Baggio}\ \bibnamefont {Saitovitch}}, \ and\ \bibinfo
  {editor} {\bibfnamefont {H.}~\bibnamefont {Petrilli}}}\ (\bibinfo
  {publisher} {Springer Berlin Heidelberg},\ \bibinfo {address} {Berlin,
  Heidelberg},\ \bibinfo {year} {2008})\ pp.\ \bibinfo {pages}
  {495--499}\BibitemShut {NoStop}%
\bibitem [{\citenamefont {Klose}\ \emph {et~al.}(2019)\citenamefont {Klose}
  \emph {et~al.}}]{Klose2019}%
  \BibitemOpen
  \bibfield  {author} {\bibinfo {author} {\bibfnamefont {A.}~\bibnamefont
  {Klose}} \emph {et~al.},\ }\bibfield  {title} {\enquote {\bibinfo {title}
  {Ground-state electromagnetic moments of $^{37}\mathrm{Ca}$},}\ }\href
  {\doibase 10.1103/PhysRevC.99.061301} {\bibfield  {journal} {\bibinfo
  {journal} {Physical Review C}\ }\textbf {\bibinfo {volume} {99}},\ \bibinfo
  {pages} {061301} (\bibinfo {year} {2019})}\BibitemShut {NoStop}%
\bibitem [{\citenamefont {Pastore}\ \emph {et~al.}(2013)\citenamefont
  {Pastore}, \citenamefont {Pieper}, \citenamefont {Schiavilla},\ and\
  \citenamefont {Wiringa}}]{Pastore2013}%
  \BibitemOpen
  \bibfield  {author} {\bibinfo {author} {\bibfnamefont {S.}~\bibnamefont
  {Pastore}}, \bibinfo {author} {\bibfnamefont {S.~C.}\ \bibnamefont {Pieper}},
  \bibinfo {author} {\bibfnamefont {R.}~\bibnamefont {Schiavilla}}, \ and\
  \bibinfo {author} {\bibfnamefont {R.~B.}\ \bibnamefont {Wiringa}},\
  }\bibfield  {title} {\enquote {\bibinfo {title} {{Quantum Monte Carlo
  calculations of electromagnetic moments and transitions in A$\leq$ 9 nuclei
  with meson-exchange currents derived from chiral effective field theory}},}\
  }\href {\doibase 10.1103/PhysRevC.87.035503} {\bibfield  {journal} {\bibinfo
  {journal} {Physical Review C}\ }\textbf {\bibinfo {volume} {87}},\ \bibinfo
  {pages} {035503} (\bibinfo {year} {2013})}\BibitemShut {NoStop}%
\bibitem [{\citenamefont {Garcia~Ruiz}\ \emph {et~al.}(2015)\citenamefont
  {Garcia~Ruiz} \emph {et~al.}}]{GarciaRuiz2015}%
  \BibitemOpen
  \bibfield  {author} {\bibinfo {author} {\bibfnamefont {R.~F.}\ \bibnamefont
  {Garcia~Ruiz}} \emph {et~al.},\ }\bibfield  {title} {\enquote {\bibinfo
  {title} {Ground-state electromagnetic moments of calcium isotopes},}\ }\href
  {\doibase 10.1103/PhysRevC.91.041304} {\bibfield  {journal} {\bibinfo
  {journal} {Physical Review C}\ }\textbf {\bibinfo {volume} {91}},\ \bibinfo
  {pages} {041304} (\bibinfo {year} {2015})}\BibitemShut {NoStop}%
\end{thebibliography}%

\end{document}